\documentclass[10pt,a4paper,journal,english]{IEEEtran}
\newif\ifdraft
\draftfalse

\usepackage[T1]{fontenc}
\usepackage[utf8]{inputenc}

\usepackage{babel}
\usepackage[singlelinecheck=true,subrefformat=parens,skip=6pt]{subcaption}
\usepackage{graphicx}
\usepackage{rotating}
\usepackage{multirow}
\usepackage{amscd}
\usepackage[centertags,reqno]{amsmath}
\usepackage{amssymb}
\usepackage{theorem}
\usepackage{paralist}
\usepackage{verbatim}
\usepackage{balance}
\usepackage{cite}
\usepackage[draft]{hyperref}
\usepackage[multiple]{footmisc}

\allowdisplaybreaks

\DeclareMathOperator*{\argmax}{arg\,max}

\newcommand{\floor}[1]{\ensuremath{\left\lfloor#1\right\rfloor}}

\newcommand{\prob}[1]{\ensuremath{\Pr\left\{#1\right\}}}
\renewcommand{\vec}[1]{\ensuremath{\mathbf{#1}}}
\newcommand{\sw}[2]{\ensuremath{_{#1}^{#2}}}
\newcommand{\field}[1]{\ensuremath{\mathbb{#1}}}

\newcommand{\sequence}[3]{\ensuremath{(#1_{#2}, \ldots, #1_{#3-1})}}
\newcommand{\seq}[2]{\ensuremath{\vec{#1}\sw{0}{#2}}}

\newcommand{\Pri}{\ensuremath{P_\mathrm{i}}}
\newcommand{\Prd}{\ensuremath{P_\mathrm{d}}}
\newcommand{\Prs}{\ensuremath{P_\mathrm{s}}}
\newcommand{\Prt}{\ensuremath{P_\mathrm{t}}}
\newcommand{\Prr}{\ensuremath{P_\mathrm{r}}}
\newcommand{\Pre}{\ensuremath{P_\mathrm{e}}}

\newcommand{\dlmin}{\ensuremath{d_{l_\mathrm{min}}}}

\newlength\figwidth
\newcommand{\insertfig}[3]{%
   \begin{figure}[!t]
      \centering
      \includegraphics[width=\figwidth]{#2}
      \caption{#3}
      \label{#1}
   \end{figure}}

\newcommand{\inserttab}[4]{%
   \begin{table}[!t]
      \begin{minipage}{\columnwidth}
         \caption{#3}
         \label{#1}
         \centering
         \renewcommand{\arraystretch}{1.15}
         \begin{tabular}{#2}
            \hline
            #4
            \hline
         \end{tabular}
      \end{minipage}
   \end{table}}

\newcommand{\inserttabd}[4]{%
   \begin{table*}[!t]
      \begin{minipage}{\textwidth}
         \caption{#3}
         \label{#1}
         \centering
         \renewcommand{\arraystretch}{1.15}
         \begin{tabular}{#2}
            \hline
            #4
            \hline
         \end{tabular}
      \end{minipage}
   \end{table*}}

\title{Time-Varying Block Codes for Synchronization Errors:
   MAP Decoder and Practical Issues}

\author{%
   Johann~A.~Briffa,
   Victor~Buttigieg,~\IEEEmembership{Member,~IEEE,}
   and~Stephan~Wesemeyer%
   \thanks{%
      Manuscript submitted February 27, 2014;
      accepted May 29, 2014 for publication in
      the IET Journal of Engineering.
      This research has been carried out using computational facilities
      procured through the European Regional Development Fund, Project
      ERDF-080.
      }%
   \thanks{%
      J.~A.~Briffa and S.~Wesemeyer are with the
      Dept.\ of Computing, University of Surrey, Guildford GU2 7XH, England.
      Email: j.briffa@surrey.ac.uk;
      s.wesemeyer@surrey.ac.uk
      }%
   \thanks{%
      V.~Buttigieg is with the Dept.\ of Comm.\ \& Computer Engineering,
      University of Malta, Msida MSD 2080, Malta.
      Email: victor.buttigieg@um.edu.mt
      }%
   }

\begin{document}

\maketitle

\begin{abstract}
In this paper we consider Time-Varying Block (TVB) codes, which generalize
a number of previous synchronization error-correcting codes.
We also consider various practical issues related to MAP decoding of
these codes.
Specifically, we give an expression for the expected distribution of drift
between transmitter and receiver due to synchronization errors.
We determine an appropriate choice for state space limits based on the drift
probability distribution.
In turn, we obtain an expression for the decoder complexity under given
channel conditions in terms of the state space limits used.
For a given state space, we also give a number of optimizations that reduce
the algorithm complexity with no further loss of decoder performance.
We also show how the MAP decoder can be used in the absence of known frame
boundaries, and demonstrate that an appropriate choice of decoder parameters
allows the decoder to approach the performance when frame boundaries are
known, at the expense of some increase in complexity.
Finally, we express some existing constructions as TVB codes, comparing
performance with published results, and showing that improved performance
is possible by taking advantage of the flexibility of TVB codes.
\end{abstract}

\begin{IEEEkeywords}
  Insertion-Deletion Correction,
  MAP Decoder,
  Forward-Backward Algorithm
\end{IEEEkeywords}


\section{Introduction}
\label{sec:intro}

Most error-control systems are designed to detect and/or correct substitution
errors, where individual symbols of the received sequence have been substituted
while maintaining synchronization with the transmitted sequence.
Some channels, however, also experience \emph{synchronization errors},
where symbols may additionally be deleted from or inserted into the received
sequence.
It has long been recognized that codes can be designed specifically for
synchronization error correction \cite{sell62, leve66}.
Except for the less-known work by Gallager \cite{gallager61}, for a long time
only some short block codes were known.
This changed when Davey and MacKay \cite{dm01ids} proposed a concatenated
scheme combining an outer LDPC code with good error-correction capability with
an inner code whose aim is to correct synchronization errors.
Ratzer \cite{ratzer05telecom} took a different approach, using short marker
sequences inserted in binary LDPC codewords; a similar approach was used by
Wang \emph{et al.} \cite{wfd2011transcomm}.
Yet another approach extends the state space of convolutional codes to
allow correction of synchronization errors \cite{sfs2004africon, spm2008sitis,
mansour2010jsac}.
The problem of convolutional code design for synchronization error channels
has been considered in \cite{mans10}.
More recently, this approach has been applied successfully to turbo codes
\cite{mans12}.
The renewed increase in interest is mainly due to new applications requiring
such codes.
A recent survey can be found in \cite{mbt2010commsurveys}.

We have in previous papers extended the work of Davey-MacKay, proposing a
\emph{maximum a posteriori} (MAP) decoder \cite{bsw10icc}, improved code
designs \cite{bs08isita,bb11isit}, as well as a parallel implementation
of the MAP decoder resulting in speedups of up to two orders of magnitude
\cite{briffa13jcomml}.
However, these papers were restricted to the case where the frame boundaries
were known by the decoder.
While Davey and MacKay showed that the frame boundaries could be accurately
determined for their bit-level decoder and code construction \cite{dm01ids},
it has not been shown whether this property extends to our MAP decoder and
improved constructions.

In this paper we define Time-Varying Block (TVB) codes in terms of the
encoding used in \cite{briffa13jcomml}, and show that TVB codes represent a
new class of codes which generalizes a number of previous synchronization
error-correcting codes.
We use the MAP decoder of \cite{briffa13jcomml} for these codes, showing how it
can be used in an iterative scheme with an outer code.
We also consider a number of important issues related to any practical
implementation of the MAP decoder.
Specifically, we give an expression for the expected distribution of drift
between transmitter and receiver due to synchronization errors.
We determine an appropriate choice for state space limits based on the drift
probability distribution.
In turn, we obtain an expression for the decoder complexity under given
channel conditions in terms of the state space limits used.
For a given state space, we also give a number of optimizations that reduce
the algorithm complexity with no further loss of decoder performance.
We also show how the MAP decoder can be used for \emph{stream decoding},
where the boundaries of the received frames are not known \emph{a priori}.
In doing so we demonstrate how an appropriate choice of decoder parameters
allows stream decoding to approach the performance when frame boundaries
are known, at the expense of some increase in complexity.
We express some previously published codes as TVB codes, comparing performance
with published results, and showing that the greater flexibility of TVB
codes permits the creation of improved codes.

In the following, we start with definitions in Section~\ref{sec:background}
and summaries of results from earlier work.
The applicable design criteria for TVB codes are considered in
Section~\ref{sec:tvbcodes}, together with the representation of previously
published codes as TVB codes.
The appropriate choice for state space limits is given in
Section~\ref{sec:statespace}, followed by expressions for the decoder
complexity in Section~\ref{sec:complexity}.
MAP decoder optimizations are given in Section~\ref{sec:optimization} and
the changes necessary for stream decoding in Section~\ref{sec:streamdecoding}.
Finally, practical results are given in Section~\ref{sec:results}.


\section{Background}
\label{sec:background}


\subsection{TVB Codes}
\label{sec:system_description}

Consider the encoding defined in \cite{briffa13jcomml}, used there to simplify
the representation of the inner code of \cite{dm01ids}.
We observe that this encoding generalizes a number of additional previous
schemes, including the inner codes of \cite{ratzer05telecom, wfd2011transcomm,
bb11isit} (c.f.\ Section~\ref{sec:system_generalization}).
We define a TVB code in terms of this encoding by the sequence
$\mathcal{C}=\sequence{C}{0}{N}$, which consists of the constituent encodings
$C_i: \mathbb{F}_q \hookrightarrow \mathbb{F}_2^n$ for $i=0,\ldots,N-1$,
where $n,q,N \in \mathbb{N}$, $2^n \geq q$, and $\hookrightarrow$ denotes
an injective mapping.
Two constituent encodings $C_i,C_j$ are said to be \emph{equal} if
$C_i(D) \in C_j \quad\forall D$.
For a given TVB code, the set of \emph{unique} constituent encodings is that
set where no two constituent encodings are equal; the cardinality of this set,
denoted by $M \leq N$, is called the order of the code.
Note that unique constituent encodings may still have some common codewords.
We denoted a TVB code by the tuple $(n,q,M)$.
We restrict ourselves to binary TVB codes, where codewords are sequences of
bits; the extension to the non-binary case is trivial.

For any sequence $\vec{z}$, denote arbitrary subsequences as
$\vec{z}\sw{a}{b} = \sequence{z}{a}{b}$, where $\vec{z}\sw{a}{a} = ()$ is an
empty sequence.
Given a message $\vec{D}\sw{0}{N} = \sequence{D}{0}{N}$, each $C_i$ maps
the $q$-ary message symbol $D_i \in \mathbb{F}_q$ to codeword $C_i(D_i)$
of length $n$.
That is, $\vec{D}\sw{0}{N}$ is encoded as
$\vec{X}\sw{0}{nN} = C_0(D_0) \| \cdots \| C_{N-1}(D_{N-1})$,
where $\vec{y}\|\vec{z}$ is the juxtaposition of $\vec{y}$ and $\vec{z}$.
Each $q$-ary symbol is encoded independently of previous inputs, and different
codebooks may be used for each input symbol.
This time-variation offers no advantage on a fully synchronized channel.
However, in the presence of synchronization errors, the differences between
neighbouring codebooks provide useful information to the decoder to recover
synchronization.



In practice a TVB code is suitable as an inner code to correct
synchronization errors in a serially concatenated construction.
A conventional outer code corrects residual substitution errors.
In such a scheme, the inner code's MAP decoder \emph{a posteriori}
probabilities (APPs) are used to initialize the outer decoder.
The concatenated code can be iteratively decoded, in which case the prior
symbol probabilities of the inner decoder are set using extrinsic information
from the previous pass of the outer decoder.


\subsection{Channel Model}
\label{sec:channel}

We consider the Binary Substitution, Insertion, and Deletion (BSID) channel,
an abstract random channel with unbounded synchronization and substitution
errors, originally presented in \cite{bahl75} and more recently used in
\cite{dm01ids, ratzer05telecom, bs08isita, bsw10icc, bb11isit} and others.
At \emph{time} $t$, one bit enters the channel, and one of three events
may happen:
insertion with probability $\Pri$ where a random bit is output;
deletion with probability $\Prd$ where the input is discarded; or
transmission with probability $\Prt=1-\Pri-\Prd$.
A substitution occurs in a transmitted bit with probability $\Prs$.
After an insertion the channel remains at time $t$ and is subject to
the same events again, otherwise it proceeds to time $t+1$, ready for
another input bit.

We define the \emph{drift} $S_t$ at time $t$ as the difference between
the number of received bits and the number of transmitted bits before the
events of time $t$ are considered.
As in \cite{dm01ids}, the channel can be seen as a Markov process with the
state being the drift $S_t$.
It is helpful to see the sequence of states as a trellis diagram, observing
that there may be more than one way to achieve each state transition.
Also note that the state space is unlimited for positive drifts but
limited for negative drifts.
Specifically, $S_t$ may take any positive value for $t>0$, though with
decreasing probability as the value increases.
On the other hand, $S_t \geq -t$ where the lower limit corresponds to receiving
the null sequence.


\subsection{The MAP Decoder}
\label{sec:mapdecoder}

We summarize here the MAP decoder of \cite{briffa13jcomml}; this is the same
as the MAP decoder of \cite{bsw10icc} with a trivial modification to work
with the notation of TVB codes.
The decoder uses the standard forward-backward algorithm for hidden Markov
models.
We assume a message sequence $\vec{D}\sw{0}{N}$, encoded using a $(n,q,M)$
TVB code to the sequence $\seq{X}{\tau}$, where $\tau = nN$.
The sequence $\seq{X}{\tau}$ is transmitted over the BSID channel, resulting
in the received sequence $\seq{Y}{\rho}$, where in general $\rho$ is not
equal to $\tau$.
To avoid ambiguity, we refer to the message sequence as a \emph{block}
of size $N$ and the encoded sequence as a \emph{frame} of size $\tau$.
We calculate the APP $L_i(D)$ of having encoded symbol $D \in \field{F}_q$ in
position $i$ for $0 \leq i < N$, given the entire received sequence, using
\begin{align}
   \label{eqn:L}
   L_i(D) & = \frac{1}{\lambda_N(\rho-\tau)} \sum_{m',m} \sigma_i(m',m,D)
   \text{,}\\
   \text{where~}
   \label{eqn:lambda}
   \lambda_i(m) &= \alpha_i(m) \beta_i(m)
   \text{,}\\
   \label{eqn:sigma}
   \sigma_i(m',m,D) &= \alpha_i(m') \gamma_i(m',m,D) \beta_{i+1}(m)
   \text{,}
\end{align}
and $\alpha_i(m)$, $\beta_i(m)$, and $\gamma_i(m',m,D)$ are the forward,
backward, and state transition metrics respectively.
Note that strictly, the above metrics depend on $\seq{Y}{\rho}$, but for
brevity we do not indicate this dependence in the notation.
The summation in \eqref{eqn:L} is taken over the combination of $m',m$,
being respectively the drift before and after the symbol at index $i$.
The forward and backward metrics are obtained recursively using
\begin{align}
   \label{eqn:alpha}
   \alpha_i(m) &= \sum_{m',D} \alpha_{i-1}(m') \gamma_{i-1}(m',m,D)\text{,}\\
   \label{eqn:beta}
   \text{and~}
   \beta_i(m) &= \sum_{m',D} \beta_{i+1}(m') \gamma_i(m,m',D)\text{.}
\end{align}
Initial conditions for known frame boundaries are given by
\begin{equation*}
\alpha_0(m) =
  \begin{cases}
     1  & \text{if $m = 0$} \\
     0  & \text{otherwise,}
  \end{cases}
\text{and~}
\beta_N(m) =
  \begin{cases}
     1  & \text{if $m = \rho-\tau$} \\
     0  & \text{otherwise.}
  \end{cases}
\end{equation*}
Finally, the state transition metric is defined as
\begin{equation}
\label{eqn:gamma}
\gamma_i(m',m,D) =
   \prob{ D_i = D }
   R( \vec{Y}\sw{ni+m'}{n(i+1)+m} \mid C_i(D) )
\end{equation}
where $C_i(D)$ is the $n$-bit sequence encoding $D$ and
$R( \vec{\dot{y}} | \vec{x} )$ is the probability of receiving a sequence
$\vec{\dot{y}}$ given that $\vec{x}$ was sent through the channel (we refer
to this as the receiver metric).
The \emph{a priori} probability $\prob{ D_i=D }$ is determined by the source
statistics, which we generally assume to be equiprobable so that
$\prob{ D_i=D } = 1/q$.
In iterative decoding, the prior probabilities are set using extrinsic
information from the previous pass of an outer decoder, as explained in
Section~\ref{sec:system_description}.
The receiver metric is obtained by calculating the forward recursion
\begin{equation}
   \label{eqn:alpha_dot}
   \dot{\alpha}_t(m)
      = \sum_{m'} \dot{\alpha}_{t-1}(m')
      \cdot
      Q\left( \vec{\dot{y}}\sw{t-1+m'}{t+m}| x_{t-1} \right)
      \text{,}
\end{equation}
where for brevity we do not show the dependence on $\vec{\dot{y}}$ and
$\vec{x}$, and $Q(\vec{y}|x)$ can be directly computed from $\vec{y},x$
and the channel parameters:
\begin{align*}
\begin{split}
   &Q( \vec{y} | x) =
   \begin{cases}
      \Prd
         & \text{if $\mu = 0$} \\
      \left(\frac{\Pri}{2}\right)^{\mu-1}
      \left( \Prt\Prs + \frac{1}{2} \Pri \Prd \right)
         & \text{if $\mu > 0, y_{\mu-1} \neq x$} \\
      \left(\frac{\Pri}{2}\right)^{\mu-1}
      \left( \Prt\bar\Prs + \frac{1}{2} \Pri \Prd \right)
         & \text{if $\mu > 0, y_{\mu-1} = x$}, \\
   \end{cases}
\end{split}
\end{align*}
where $\mu$ is the length of $\vec{y}$ and $\bar\Prs = 1-\Prs$.
The required value of the receiver metric is given by
$R( \vec{\dot{y}} | \vec{x} ) = \dot{\alpha}_{n}(\dot{\mu}-n)$,
where $\dot{\mu}$ is the length of $\vec{\dot{y}}$, $n$ is the length of
$\vec{x}$.

As in \cite{briffa13jcomml}, the $\alpha$, $\beta$, and $\dot\alpha$
metrics are normalized as they are computed to avoid exceeding the limits
of floating-point representation.
We also assume that \eqref{eqn:alpha_dot} is computed at single
precision\footnote{%
We refer to 32-bit floating point as single precision, and 64-bit floating
point as double precision.},
while the remaining equations use double precision.


\section{TVB Code Design}
\label{sec:tvbcodes}

\subsection{Construction Criteria}
\label{sec:construction}

In any error-correcting scheme, the decoder's objective is to minimize the
probability of decoding error (at the bit or codeword level depending on
the application).
If the channel does not introduce synchronization errors, this optimization
may be performed independently of previous or subsequent codewords.
Hence the performance of the code depends exclusively on its distance
properties.
In particular, the performance of the code at low channel error rate is
dominated by the code's minimum Hamming distance.
At any channel error rate the performance is determined by the code's distance
spectrum \cite{pere96,ferr03}.
Thus when designing codes for substitution error channels, either the minimum
Hamming distance or the more complete distance spectrum needs to be optimized
for the given code parameters.

In the case of the BSID channel, and other channels that allow synchronization
errors, a similar behaviour is observed if the codeword boundaries are known,
only this time the Levenshtein distance \cite{leve66} replaces the Hamming one.
Recall that the Levenshtein distance gives the minimum number of edits
(insertions, deletions, or substitutions) that will change one codeword
into another.

For the BSID channel an upper bound for the probability of decoding a codeword
in error was given in \cite{bb11isit}, assuming codeword boundaries are known.
For $\Pri,\Prd,\Prs \ll 1$, the bound of \cite[(9)]{bb11isit} is
dominated by the number of correctable errors, $t$.
Now, for a code with minimum Levenshtein distance $\dlmin$, it can be shown
that $t = \floor{\frac{\dlmin-1}{2}}$ \cite{leve66}.
Hence designing TVB codes with constituent encodings having large $\dlmin$
will result in the greatest improvement to the code's performance at low
channel error rates.

However, in general the codeword boundaries are not known and need to be
estimated by the decoder.
Therefore decoding a given codeword on synchronization error channels
depends not only on the current received word, but also on previous and
subsequent ones.
This means that the performance of a TVB code depends not only on the distance
properties of constituent encodings considered separately, but also on the
relationship between constituent encodings.
This effect becomes more significant under poorer channel conditions, where
the drift can easily exceed the length of a codeword.
Unfortunately, the required relationship between constituent encodings for
optimal performance over the BSID channel is still an open problem.
What is known is that the diversity created by a sequence of different
encodings helps the decoder estimate the drift within a codeword length,
improving performance at higher channel error rates \cite{bb11isit}.

\subsection{Representation of Previous Schemes as TVB Codes}
\label{sec:system_generalization}

TVB codes generalize a number of existing synchronization error-correcting
codes.
The flexibility of the generalization allows the creation of improved codes
at the same size and rate, as we shall show.

%
Consider first the sparse inner codes with a distributed marker
sequence\footnote{This was originally referred to as a watermark sequence.} of
the Davey-MacKay construction \cite{dm01ids}.
It is clear that the sparse code is a fixed encoding
$C': \mathbb{F}_q \hookrightarrow \mathbb{F}_2^n$;
these codewords are then added to a distributed marker sequence
$\vec{w}_i$ of length $n$, specific for each codeword index $i$.
Thus we can write $C_i(D_i) = C'(D_i) + \vec{w}_i$ to represent the
inner codes of \cite{dm01ids} as TVB codes.
The equivalence of this mapping to the inner code of \cite{dm01ids} has
also been shown in \cite{briffa13jcomml}.
The distributed marker serves the same function as the use of different
encodings in TVB codes.
The decoder of \cite{dm01ids} tracks the marker sequence directly, treating
the additive encoded message sequence as substitution errors.
Therefore, to corrupt the marker sequence as little as possible, the inner
code used is sparse.
The sparseness results in a low $\dlmin$, making it harder for the decoder
to distinguish between the various codewords, and leads to relatively poor
performance at low channel error rates.

%
The codes of \cite{bb11isit} can similarly be represented as TVB codes,
with $C'$ corresponding to the Synchronization and Error Correcting (SEC)
code and $\vec{w}_i$ corresponding to the Allowed Modification Vectors (AMVs).
SEC codes are designed with a large $\dlmin$ for good performance at low
channel error rates.
For such channels this code can perform much better than the sparse code
of \cite{dm01ids}.
AMVs are chosen such that when added to the SEC code the resulting code's
$\dlmin$ does not change.
Clearly, the AMVs serve the same function as the use of different encodings
in TVB codes.
In contrast to a random distributed marker sequence, the use of AMVs does
not compromise the performance of the underlying SEC code at low channel
error rates.
In general, however, the Levenshtein distance spectrum is altered.
The separate constituent encodings in TVB codes give greater design freedom
than SEC codes with AMVs and also allows the design of constituent encodings
that maintain the required optimized Levenshtein distance spectrum.

%
The marker codes given by Ratzer \cite{ratzer05telecom} can also be cast as
TVB codes by letting each possible sequence of data bits (between markers)
be represented by a $q$-ary symbol.
For example, consider a marker code with 3 marker bits inserted after every
9 data bits, where the 3-bit marker is randomly chosen between the sequences
$001$ and $110$.
This can be represented as a $(12,512,2)$ TVB code, where encoding $C_0$
consists of all possible 9-bit sequences appended with $001$, and $C_1$
consists of all possible 9-bit sequences appended with $110$.
Like the sparse codes of \cite{dm01ids}, these marker codes suffer from a low
$\dlmin$, leading to relatively poor performance at low channel error rates.
On the other hand, the fixed marker bits improve the determination of codeword
boundaries, and the random use of different marker bits creates the necessary
diversity to improve performance in poorer channel conditions.

%
To illustrate the difference in performance between the various designs,
consider encodings of size $(n,q)=(7,8)$ with $N=666$
(same size as codes C and H in \cite{dm01ids}).
A $(7,8,4)$ TVB code where each constituent code has the best possible
Levenshtein distance spectrum with $\dlmin=3$, found through an exhaustive
search, is given in Table~\ref{tab:exampletvb}.
\inserttab{tab:exampletvb}{cccc}
   {A $(7,8,4)$ TVB code $\mathcal{C} = (C_0,\ldots,C_3$) with $\dlmin=3$.}{
   \emph{$C_0$} &
   \emph{$C_1$} &
   \emph{$C_2$} &
   \emph{$C_3$} \\
   \hline
   0000000 & 0000000 & 0000011 & 0000000 \\
   0000111 & 0000111 & 0001100 & 0001111 \\
   0011001 & 0011110 & 0011111 & 0101001 \\
   0110110 & 0110101 & 0101010 & 0110110 \\
   1001010 & 1001001 & 1011001 & 1000011 \\
   1100001 & 1100110 & 1100000 & 1001100 \\
   1111000 & 1111000 & 1100111 & 1110000 \\
   1111111 & 1111111 & 1111110 & 1111111 \\
   }
In Fig.~\ref{fig:inner} we compare this TVB code with earlier constructions
from the literature at the same size.
\insertfig{fig:inner}{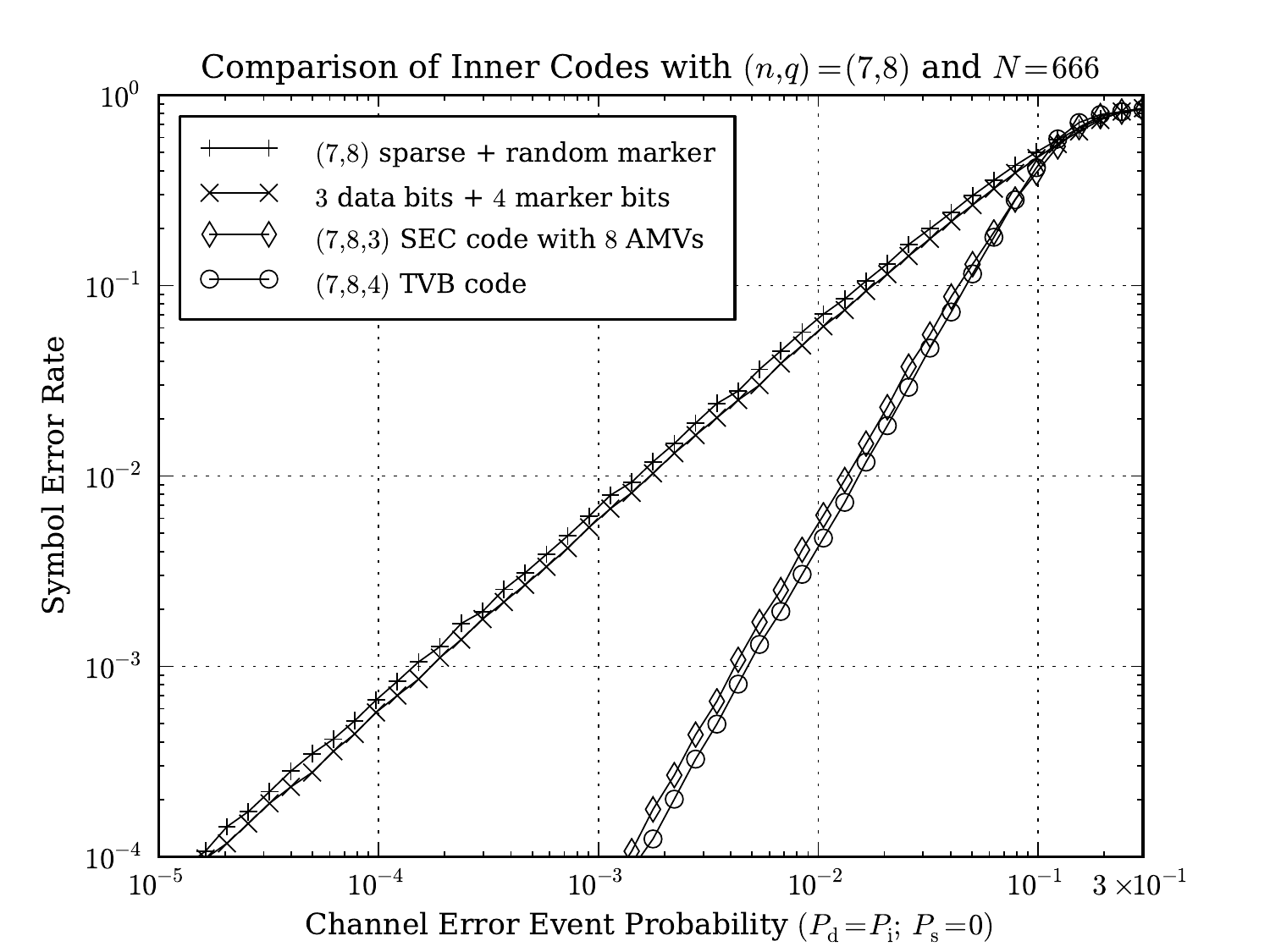}
   {Comparison of inner code designs of size $(n,q)=(7,8)$ with $N=666$:
   a sparse code with random distributed marker from \cite{dm01ids},
   a marker code with $0011/1100$ marker bits similar to \cite{ratzer05telecom},
   a SEC code with randomly-sequenced AMVs from \cite{bb11isit},
   and the TVB code of Table~\ref{tab:exampletvb}.}
Consider first the SEC code of the same size and $\dlmin$ from \cite{bb11isit},
used with 8 AMVs in a random sequence.
As expected, the TVB code performs better due to its improved Levenshtein
distance spectrum, even though both TVB and SEC codes have the same $\dlmin$.
The performance of the sparse code with random distributed marker from
\cite{dm01ids} is considerably worse, particularly at low channel error rates.
Similarly a code with three data bits and four marker bits (randomly chosen
between $0011/1100$), similar to \cite{ratzer05telecom}, also performs poorly
at low channel error rates.



\section{Appropriate Limits on State Space}
\label{sec:statespace}

The equations in Section~\ref{sec:mapdecoder} assume that summations can be
taken over the set of all possible states.
For a channel such as the one considered, the state space is unbounded for
positive drifts.
A practical implementation will have to take sums over a finite subset
of states.
In \cite{dm01ids} the state space was limited to a drift $|S_t| \le
x_\textrm{max}$, where $x_\textrm{max}$ was chosen to be `several times
larger' than the standard deviation of the synchronization drift over one
block length, assuming this takes a Gaussian distribution.
No recommendation was given for the value that should be used.

Limiting the state space is by definition sub-optimal.
However, we can arbitrarily lower the number of cases where the sub-optimal
solution is worse than the optimal one, by ensuring that only the least
likely states are omitted.
The choice of summation limits also involves a trade-off with complexity,
which has a polynomial relationship with the size of the state space
(c.f.\ Section~\ref{sec:complexity}).
Therefore, an appropriate choice of summation limits will result in the
smallest state space such that the probability of the drift being outside
that range is as low as required.
The first step to identify good summation limits is to derive an accurate
probability distribution of the state space, avoiding the Gaussian
approximation of \cite{dm01ids}.

\subsection{Drift Probability Distribution}

The drift $S_T$ after transmission of $T$ bits was stated in \cite{dm01ids}
(and shown in \cite{davey99}) to be normally distributed with zero mean and a
variance equal to $T p / (1-p)$ for the special case where $p := \Pri = \Prd$.
This distribution is asymptotically valid as $T \rightarrow \infty$.
For cases where $\Pri \neq \Prd$ or where $T$ is not large enough, this
distribution cannot be used.
This is particularly relevant for determining the summation limits of
\eqref{eqn:alpha_dot} where the sequence length $n$ is not large.
An exact expression for the probability distribution of $S_T$ is given by
\ifdraft
   \begin{equation}
   \Phi_T(m) = \prob{S_T = m}
      = \Prt^{T} \Pri^{m} \sum_{j=j_0}^{T}
         \binom{T}{j} \binom{T + m + j - 1}{m + j}
         \left[ \frac{\Pri \Prd}{\Prt} \right] ^j
         \text{,}
         \label{eqn:driftpdf}
   \end{equation}
\else
   \begin{align}
   & \Phi_T(m) = \prob{S_T = m} \nonumber\\
   & \quad = \Prt^{T} \Pri^{m} \sum_{j=j_0}^{T}
         \binom{T}{j} \binom{T + m + j - 1}{m + j}
         \left[ \frac{\Pri \Prd}{\Prt} \right] ^j
         \text{,}
         \label{eqn:driftpdf}
   \end{align}
\fi
where $j_0=\max(-m,0)$.
Observe that for a drift of $m$ bits, we need $m$ insertion events more than
we have deletion events.
Over a sequence of $T$ bits, for $j$ deletion events, this means $m+j$
insertion events and $T-j$ transmission events.
The probability of this is
$\Pri^{m+j} \Prd^{j}\Prt^{T-j} =
   \Prt^{T} \Pri^{m}\left[ \Pri \Prd / \Prt \right] ^j$.
We get \eqref{eqn:driftpdf} by adding all different combinations of these
events, and summing over $j$, noting that we cannot have fewer than $0$
events of any type.
Specifically, the number of combinations for $j$ deletions in $T$ transmitted
bits is given by $\binom{T}{j}$.
The number of combinations for $m+j$ insertions is given by
$\binom{T + m + j - 1}{m + j}$, as the $m+j$ insertion events create an
additional $m+j$ opportunities for insertion.

\subsection{Avoiding Numerical Issues}
\label{sec:driftpdf_numerical}

In a practical implementation, computing the drift probability
\eqref{eqn:driftpdf} requires a few special considerations.
Practical codes from the literature have codeword size $n$ in the range
5--12 bits and number of codewords $N$ up to 1000, for a frame length $nN$ of
about 4000--6000 bits.
These codes are designed to operate under channel conditions $\Pri,\Prd$
from $10^{-3}$ to above $10^{-1}$.
Evaluating \eqref{eqn:driftpdf} under these conditions, one encounters very
large values for the two binomial coefficients and very small values for
the power term.
For example, consider evaluating \eqref{eqn:driftpdf} at $m=0$ for $T=6000$
and $\Pri=\Prd=10^{-3}$.
The two binomial coefficients have a range of up to $1.56\times10^{1804}$
(at $j=3000$) and $8.34\times10^{3609}$ (at $j=6000$) respectively.
The power term has a range of down to $1.65\times10^{-35995}$ (at $j=6000$).
This range is far beyond that representable even in double-precision
floating point.
A direct implementation of \eqref{eqn:driftpdf} will therefore result in
numerical overflow and underflow (in computing the binomial coefficients
and power term respectively) for typical frame sizes and channel conditions,
even though the summation term itself is representable.

The above numerical range problem can be avoided by combining the computation
of all terms in the summation as follows.
Observe that \eqref{eqn:driftpdf} can be rewritten as
\begin{align}
\label{eqn:phi_sum}
\Phi_T(m) &= \sum_{j=j_0}^{T} \delta_j\text{,}\\
\text{where~} \delta_j &= \Prt^{T} \Pri^{m}
      \binom{T}{j} \binom{T + m + j - 1}{m + j}
      \left[ \frac{\Pri \Prd}{\Prt} \right] ^j
      \label{eqn:deltaj}
\end{align}
and $j_0=\max(-m,0)$ as before.
In this expression, note that the summation is empty if $j_0 > T$, resulting
in zero probability.
Also, since $j \geq 0$, the first binomial coefficient is always non-zero,
while the second binomial coefficient is non-zero if $T > 0$.
Expanding the binomial coefficients using the factorial formula, we can
express the summation term recursively as
\begin{equation}
\delta_j = \delta_{j-1} \cdot
      \frac{\Pri \Prd}{\Prt} \cdot
      \frac{T + m + j - 1}{m + j} \cdot
      \frac{T - j + 1}{j}
      \text{,}
      \label{eqn:deltaj_recursive}
\end{equation}
allowing successive factors to be determined easily from previous ones.
The initial factor required is the one at $j_0$, and can be determined
from \eqref{eqn:deltaj} by expanding the binomial coefficients using the
multiplicative formula:
\begin{equation}
\delta_{j_0} = \Prt^{T} \Pri^{m}
      \prod_{i=1}^{j_0} \frac{T - j_0 - i}{i}
      \prod_{i=1}^{m+j_0} \frac{T - 1 - i}{i}
      \left[ \frac{\Pri \Prd}{\Prt} \right] ^{j_0}
      \text{.}
      \label{eqn:deltaj0}
\end{equation}
Consider the earlier example, now evaluating \eqref{eqn:phi_sum} at $m=0$ for
$T=6000$ and $\Pri=\Prd=10^{-3}$.
In this case, the initial value $\delta_{j_0} = 6.07\times10^{-6}$,
and the multiplier $\delta_j/\delta_{j-1}$ in the recursive expression
\eqref{eqn:deltaj_recursive} has its smallest value of $3.34\times10^{-10}$
at $j=6000$.
Both values are easily representable as floating point numbers.

Using \eqref{eqn:phi_sum}, numerical range issues remain when computing
$\delta_{j_0}$ for larger values of $\Pri,\Prd$, and consequently also for
storing successive values of $\delta_j$.
For example, consider evaluating \eqref{eqn:phi_sum} at $m=0$ for $T=6000$
and $\Pri=\Prd=10^{-1}$.
In this case $\delta_{j_0} = 3.47\times10^{-582}$, and one needs to accumulate
a number of $\delta_j$ values in this range to obtain the required result
$\Phi_{6000}(0) = 0.0109$.
Again, the intermediate values are beyond the range of double-precision
floating point numbers although the final result is representable.
These numerical range issues can be avoided by computing
\eqref{eqn:deltaj_recursive} and \eqref{eqn:deltaj0} using logarithms.
For the earlier example with $m=0$, $T=6000$, and $\Pri=\Prd=10^{-1}$,
we now get $\log\delta_{j_0} = -1.34\times10^3$ and the smallest value of
$\log\delta_j$ is $-1.93\times10^4$ at $j=6000$.

Finally, the required drift probability is obtained by accumulating the
exponential of the $\log\delta_j$ values using \eqref{eqn:phi_sum}.
However, the individual values of $\delta_j$ are still beyond the range of
double-precision floating point.
In practice, we have found that the use of extended-precision (80-bit) floating
point provides sufficient range.
Alternatively, the accumulation in \eqref{eqn:phi_sum} may be computed in
logarithmic domain using the property
$\log(A+B) = \log{A} + \log\left( 1 + e^{\log{B}-\log{A}} \right)$.

Note that expression \eqref{eqn:driftpdf} is valid for any
$\Pri \ge 0$, $\Prd \ge 0$, and $\Pri + \Prd < 1$.
However, the computation using logarithms cannot be applied directly when
either or both of $\Pri$ and $\Prd$ are zero.
These degenerate cases have to be handled as special cases, by first reducing
\eqref{eqn:driftpdf} and then implementing the simplified equations using
logarithms.

\subsection{Probability of Drift Outside Range}
\label{sec:drift_limits}


We want to choose lower and upper limits $m_T^{-}, m_T^{+}$ such that the
drift after transmitting a sequence of $T$ bits is outside the range
$\{ m_T^{-} \ldots m_T^{+} \}$ with an arbitrarily low probability $\Prr$:
\begin{eqnarray}
\prob{S_T < m_T^{-}} + \prob{S_T > m_T^{+}} < \Prr
\text{,}\\
\text{or equivalently:~}
\label{eqn:limit_condition}
1 - \sum_{m = m_T^{-}}^{m_T^{+}} \Phi_T(m) < \Prr
\text{.}
\end{eqnarray}
An appropriate choice of limits can be obtained iteratively as follows.
Observe that for the BSID channel $\Phi_T(m)$ is monotonically decreasing with
increasing $|m|$.
A first estimate for the limits is given by:
\begin{align}
\label{eqn:limit_lower}
m_T^{-(1)} = \max m \left|
   \Phi_T(m-1) < \frac{\Prr}{2}
      \right. \\
\label{eqn:limit_upper}
\text{and~}
m_T^{+(1)} = \min m \left|
   \Phi_T(m+1) < \frac{\Prr}{2}
      \right.
   \text{,}
\end{align}
where the number in superscript parentheses indicates the iteration count.
If these estimates satisfy \eqref{eqn:limit_condition}, we use them as our
lower and upper limits.
Otherwise these estimates are updated iteratively as follows:
\begin{align}
m_T^{-(i+1)} &=
   \begin{cases}
   m_T^{-(i)}-1 & \text{if $\Phi_T(m_T^{+(i)}+1) \le \Phi_T(m_T^{-(i)}-1)$} \\
   m_T^{-(i)}   & \text{otherwise,}
   \end{cases} \\
m_T^{+(i+1)} &=
   \begin{cases}
   m_T^{+(i)}+1 & \text{if $\Phi_T(m_T^{+(i)}+1) > \Phi_T(m_T^{-(i)}-1)$} \\
   m_T^{+(i)}   & \text{otherwise.}
   \end{cases}
\end{align}
That is, we extend the range by one in the direction of greatest gain.
The iterative process is repeated until \eqref{eqn:limit_condition}
is satisfied.
The size of the state space is given by $M_T = m_T^{+} - m_T^{-} + 1$.

\subsection{Choice of Summation Limits}
\label{sec:summation_limits}

When considering the whole frame, $T=\tau$, so that the overall size of the
state space is given by $M_\tau$.
Now the final output of the MAP decoder is calculated using \eqref{eqn:L},
which sums over all $M_\tau$ prior states $m_\tau^{-} \le m' \le m_\tau^{+}$.
For each prior state, however, only the drifts introduced by the transmission
of $n$ bits need to be considered, corresponding to a subset $M_n$ of states
$m_n^{-} \le m \le m_n^{+}$.
Similarly, the computation of \eqref{eqn:alpha} and \eqref{eqn:beta} is
required for all $M_\tau$ states $m_\tau^{-} \le m \le m_\tau^{+}$, each
involving a summation over $M_n$ prior or posterior states $m_n^{-} \le m'
\le m_n^{+}$ respectively.
Finally, the state transition metric is obtained using the forward pass
of \eqref{eqn:alpha_dot}; this is computed over a sequence of $n$ bits
for each of $M_n$ states $m_n^{-} \le m \le m_n^{+}$.
Each recursion consists of a summation over prior states $m'$; in this
case only the drifts introduced by the transmission of one bit need to be
considered, corresponding to a subset $M_1$ of prior states $m_1^{-} \le m'
\le m_1^{+}$.

Now consider that we want to limit the probability of any of these summations
not covering an actual channel event over a whole frame to, say, no more
than $\Pre$.
When computing the limits over the whole frame, $m_\tau^{\pm}$, we simply need
to set $\Prr=\Pre$.
However, when computing limits over an $n$-bit sequence, $m_n^{\pm}$,
since this summation is repeated for each of $N$ such sequences, we set
$\Prr=1-\sqrt[N]{1-\Pre} \approx \frac{\Pre}{N}$ for small $\Pre$.
Similarly, for limits over a $1$-bit sequence, $m_1^{\pm}$, we use
$\Prr=1-\sqrt[\tau]{1-\Pre} \approx \frac{\Pre}{\tau}$ for small $\Pre$.

Except in the case of stream decoding (c.f. Section~\ref{sec:streamdecoding}),
the state space limits only need to be determined once and remain valid as
long as the channel conditions do not change.
In any case, the required values of $\Phi_T(m)$ depend only on the code
parameters and channel conditions, so that a table may be pre-computed.
This makes the average complexity of determining the state space limits
negligible.

\subsection{Example}
\label{sec:limits_example}

\begin{figure}[!t]
   \centering
   \begin{subfigure}[b]{\figwidth}
      \includegraphics[width=\figwidth]{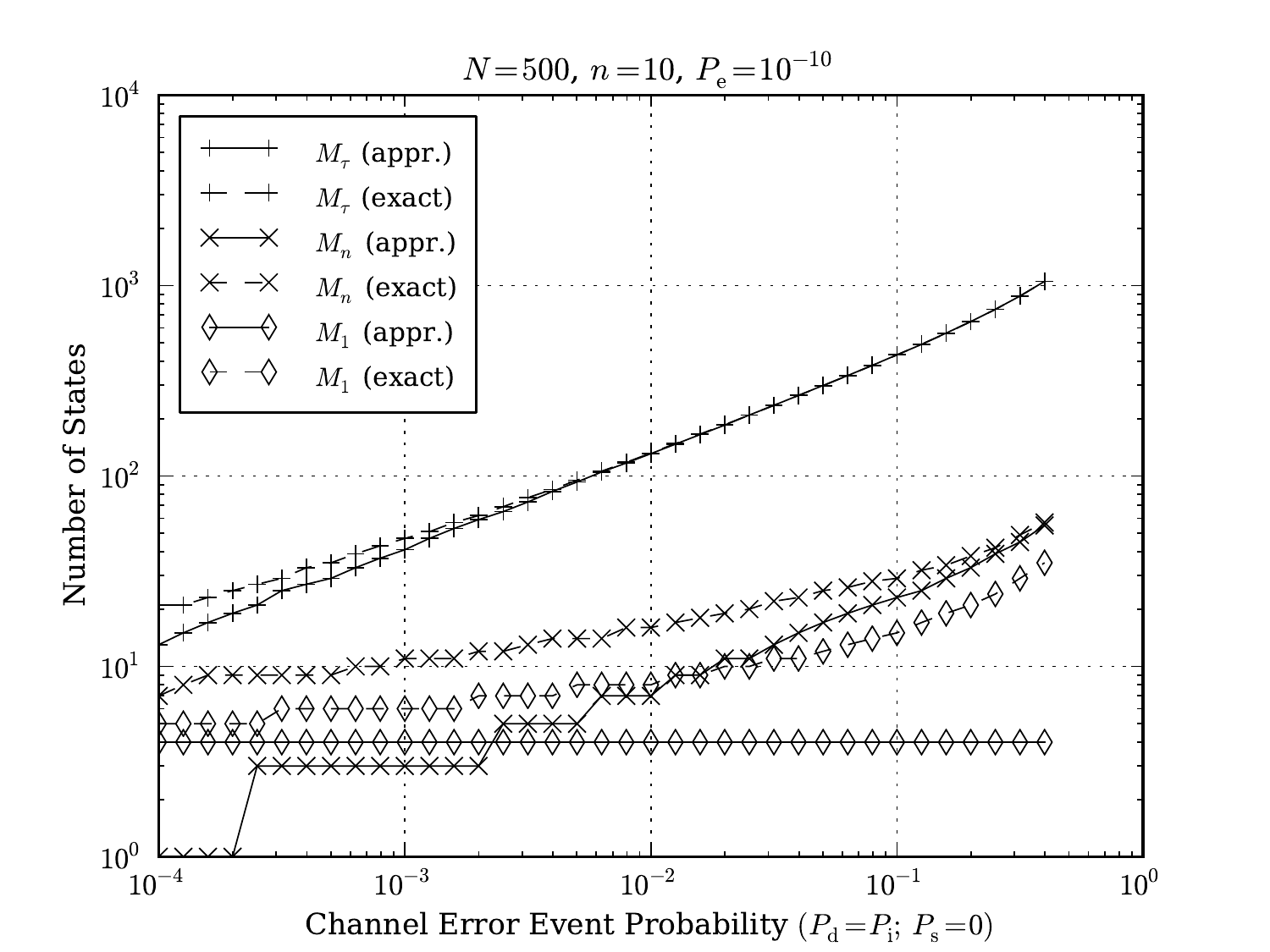}
      \caption{}
      \label{fig:statespace-size}
   \end{subfigure}
   \begin{subfigure}[b]{\figwidth}
      \includegraphics[width=\figwidth]{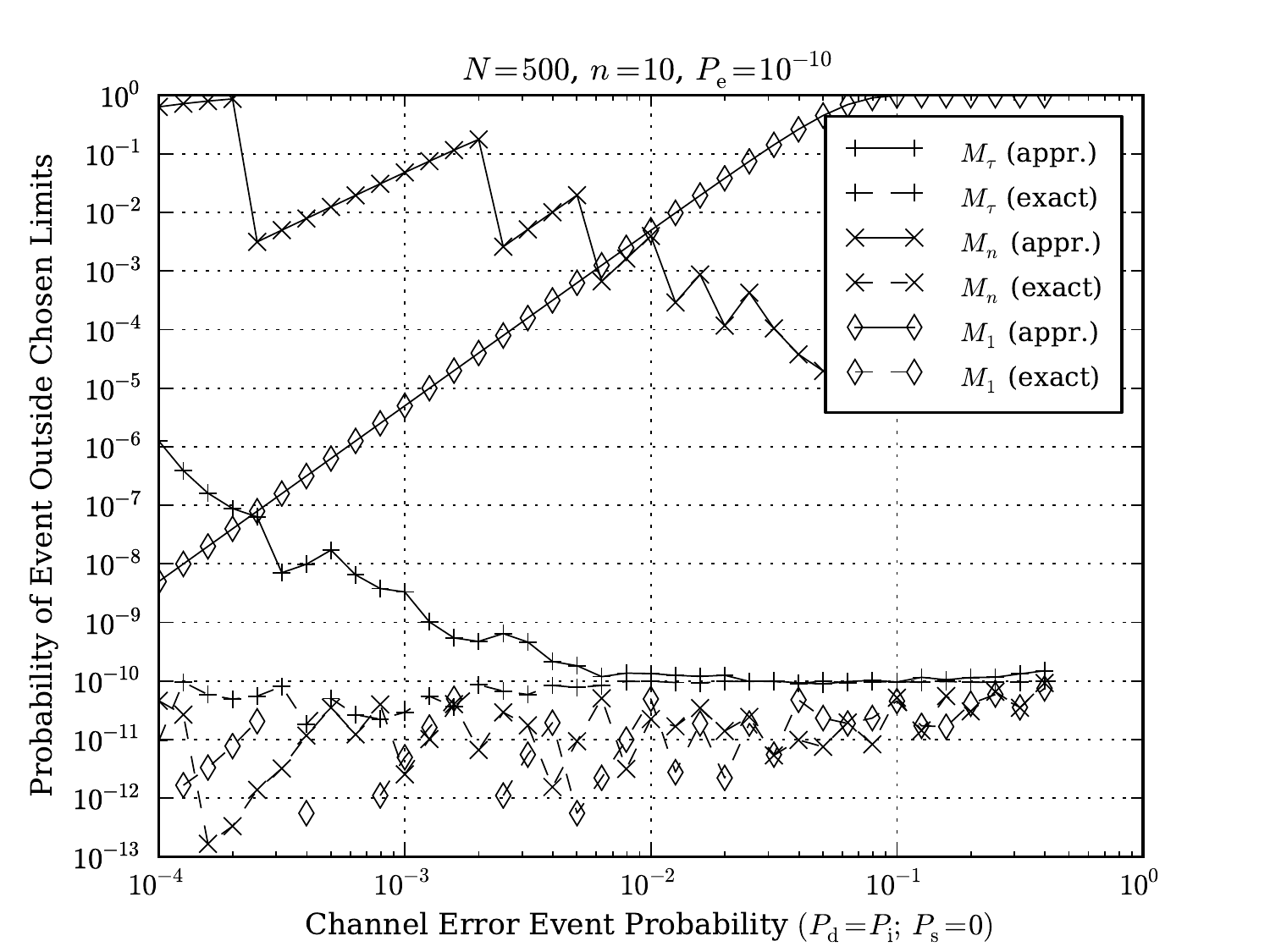}
      \caption{}
      \label{fig:statespace-cover}
   \end{subfigure}
   \caption{
      \protect\subref{fig:statespace-size} Number of states within summation limits,
      and
      \protect\subref{fig:statespace-cover} probability of encountering a channel
      event outside the chosen summation limits over a single frame,
      using the approximation of \cite{dm01ids} and our exact computation.}
   \label{fig:dminner2-750}
\end{figure}

Overestimating the required state space increases computational complexity,
while underestimating the state space often results in poor decoding
performance.
Accurate limits are particularly important for restricting the drifts
considered across each codeword.
It is therefore useful to illustrate the discrepancy between the approximate
distribution of \cite{dm01ids} and the exact expression for the distribution
of the drift.
Consider a system with typical block and codeword sizes $N=500$, $n=10$.
We plot in Fig.~\ref{fig:statespace-size} the number of states within
summation limits using the approximate and exact expressions, in each case
for $\Pre=10^{-10}$.
For $T=1$, \cite[Section~VII.A]{dm01ids} assumes a maximum of two successive
insertions; this is equivalent to setting $m_1^{+}=2$, so that $M_1=4$.
It is immediately apparent that while the approximation is very close for
large $T$ and high $\Pri=\Prd$, it quickly starts to underestimate the
required range at lower channel error rates.
As expected, the discrepancy is particularly large when considering shorter
sequences.
For $T=1$ it is not surprising that there is a large discrepancy for channels
with high error rate.

Next, we determine the probability of encountering a channel event outside the
chosen limits over a single frame, shown in Fig.~\ref{fig:statespace-cover}
for the same limits used in Fig.~\ref{fig:statespace-size}.
For the exact distribution, this probability is always below the chosen
threshold $\Pre=10^{-10}$, as expected.
For the approximation, however, the probability of exceeding the chosen limits
is higher than the threshold throughout the range considered.
At lower channel error rates, the discrepancy is significant (several orders
of magnitude) even for large $T$.
For small $T$, the probability of exceeding the chosen limits is high enough
to make the approximation useless.
For $T=1$, the artificial limit of two successive insertions of \cite{dm01ids}
means that channels with high error rate will exceed this limit with high
probability.



\section{Algorithm Complexity}
\label{sec:complexity}


\subsection{Complexity of the MAP Decoder}

As a first step towards determining the overall complexity of the
decoder, consider first the calculation of the state transition metric in
\eqref{eqn:gamma}.
This is recursively computed using the forward pass of \eqref{eqn:alpha_dot}
over a sequence of $n$ bits, for each of $M_n$ states $m$.
Each recursion consists of a summation over $M_1$ prior states $m'$ as
argued in Section~\ref{sec:summation_limits}.
The bit-level probability $Q$ can be obtained by a look-up table.
Thus the complexity for calculating a single state transition metric is
$\Theta(n M_n M_1)$.

The final output of the algorithm consists of $q$ probabilities for each of
$N$ symbols, calculated using \eqref{eqn:L}.
This equation sums over all $M_\tau$ prior states $m'$ and $M_n$ states $m$,
defining the domain for $\sigma_i(m',m,D)$.
It follows from \eqref{eqn:sigma} that the domain for $\gamma_i(m',m,D)$
is the same.
Now the computation of \eqref{eqn:sigma} is dominated by the evaluation of
$\gamma_i(m',m,D)$ in \eqref{eqn:gamma}, whose complexity is
$\Theta(n M_n M_1)$ as shown.
Considering the number of times the $\gamma$ metric is computed, the MAP
decoder has an asymptotic complexity of $\Theta(N n q M_\tau M_n^2 M_1)$.

\subsection{Complexity of the Davey-MacKay Decoder}

It would initially appear that the MAP decoder complexity is significantly
higher than that of the Davey-MacKay decoder, given as $O(N n M_\tau M_1)$
in \cite{dm01ids} for a direct implementation (using our notation).
However, the expression of the Davey-MacKay decoder seems to consider only
the complexity of the initial forward and backward passes, ignoring the
additional small forward passes needed to compute the final decoder output.

The final output of the Davey-MacKay algorithm also consists of $q$
probabilities for each of $N$ symbols.
Each of these is computed using \cite[(4)]{dm01ids}, which sums over all
possible prior and posterior states.
In a direct implementation all possible prior states need to be considered;
using our notation the number of states is $M_\tau$.
While not stated in \cite{dm01ids}, the number of posterior states that need
to be considered is $M_n$, as argued for the MAP decoder.
The computation within the summation of \cite[(4)]{dm01ids} is dominated by
the conditional probability, which is computed using a separate forward pass.
This forward pass is effectively identical to \eqref{eqn:alpha_dot}
whose complexity is $\Theta(n M_n M_1)$.
Considering the number of times the forward pass is computed, it follows
that the Davey-MacKay decoder has an overall asymptotic complexity of
$\Theta(N n q M_\tau M_n^2 M_1)$.

\subsection{Comments on Algorithm Complexity}
\label{sec:complexity_comparison}

Comparing the complexity expressions for the MAP decoder for TVB codes and
the Davey-MacKay decoder for sparse codes with a distributed marker sequence,
it follows that the asymptotic complexity for both decoders is the same.
This is consistent with experimental running times for both decoders
in \cite{bsw10icc}.

In the complexity expression note that $N$, $n$, and $q$ depend only on the
code parameters while $M_\tau$, $M_n$, and $M_1$ also depend on the channel
conditions.
For $M_1$, it was argued in \cite[Section~VII.A]{dm01ids} that it is sufficient
to consider a maximum of two successive insertions, at a minimal cost to
decoding performance.
This is equivalent to setting $m_1^{+}=2$, so that $M_1=4$;
these limits were also used in \cite{bs08isita,bsw10icc}.
However, this artificially low limit is insufficient for more advanced code
constructions, as shown in \cite{bb11isit}.
It was also argued in \cite{dm01ids} that useful speedups can be obtained
by only following paths through the trellis that pass through nodes with
probabilities above a certain threshold.
However, the choice of this threshold was not analyzed.
It is also likely that this choice would depend on the properties of the
inner code being used.


\section{Speeding Things Up}
\label{sec:optimization}

\subsection{Batch Computation of Receiver Metric}
\label{sec:batchreceiver}

In a naïve implementation, each $\gamma$ computation \eqref{eqn:gamma}
requires the computation of the receiver metric as a separate forward pass
using \eqref{eqn:alpha_dot}.
However it can be observed that for a given starting state $m'$ and symbol
$D$, the $\gamma$ metric will be computed for each end state $m$ within
the limits considered (c.f.\ equations \eqref{eqn:L}, \eqref{eqn:sigma},
\eqref{eqn:alpha}, and \eqref{eqn:beta}, where the $\gamma$ computations
are used).
In turn, this means that for a given $\vec{x}$, the receiver metric will
need to be determined for all subsequences $\dot{\vec{y}}$ within the drift
limit considered.
It is therefore sufficient to compute the forward pass \eqref{eqn:alpha_dot}
once, with the longest subsequence $\dot{\vec{y}}$ required.
In doing so, the values of the receiver metric for shorter subsequences are
obtained for free.
We call this approach \emph{batch} computation.
This effectively reduces the complexity of computing the collection of $\gamma$
metrics by a factor of $M_n$.
The asymptotic complexity of the MAP decoder is therefore reduced to
$\Theta(N n q M_\tau M_n M_1)$.


\subsection{Lattice Implementation of Receiver Metric}
\label{sec:lattice}

To compute the receiver metric, an alternative to the trellis of
\eqref{eqn:alpha_dot} is to define a recursion over a lattice as in
\cite{bahl75}.
For the computation of $R( \vec{\dot{y}} | \vec{x} )$, the required
lattice has $n+1$ rows and $\dot\mu+1$ columns.
Each horizontal path represents an insertion with probability
$\frac{1}{2}\Pri$, each vertical path is a deletion with probability $\Prd$,
while each diagonal path is a transmission with probability $\Prt\Prs$ if
the corresponding elements from $\vec{x}$ and $\vec{\dot{y}}$ are different
or $\Prt\bar\Prs$ if they are the same.
Let $F_{i,j}$ represent the lattice node in row $i$, column $j$.
Then the lattice computation in the general case is defined by the recursion
\begin{equation}
\label{eqn:F}
F_{i,j} =
   \frac{1}{2}\Pri F_{i,j-1} +
   \Prd F_{i-1,j} +
   \dot{Q}(\dot{y}_j | x_i) F_{i-1,j-1}
   \text{,}
\end{equation}
which is valid for $i<n$, and where $\dot{Q}(y|x)$ can be directly computed
from $y,x$ and the channel parameters:
\begin{equation}
\begin{split}
   &\dot{Q}( y | x) =
   \begin{cases}
      \Prt\Prs      & \text{if $y \neq x$} \\
      \Prt\bar\Prs  & \text{if $y = x$}\text{.} \\
   \end{cases}
\end{split}
\end{equation}
Initial conditions are given by
\begin{equation}
\begin{split}
   &F_{i,j} =
   \begin{cases}
      1 & \text{if $i=0$, $j=0$} \\
      0 & \text{if $i<0$ or $j<0$.} \\
   \end{cases}
\end{split}
\end{equation}
The last row is computed differently as the channel model does not allow
the last event to be an insertion.
In this case, when $i=n$, the lattice computation is defined by
\begin{equation}
\label{eqn:F_lastrow}
F_{n,j} =
   \Prd F_{n-1,j} +
   \dot{Q}(\dot{y}_j | x_n) F_{n-1,j-1}
   \text{.}
\end{equation}
Finally, the required receiver metric is obtained from this computation as
$R( \vec{\dot{y}} | \vec{x} ) = F_{n,\dot{\mu}}$.
The calculation of a single run through the lattice requires a number of
computations proportional to the number of nodes in the lattice.
Now for the transmitted sequence of $n$ bits considered, the number of rows
will always be $n$ while the number of columns is at most $n + m_n^{+}$.
The complexity of a direct implementation of this algorithm is therefore
$\Theta(n [n + m_n^{+}])$.

It has been argued in Section~\ref{sec:batchreceiver} that for a given
$\vec{x}$, the receiver metric $R( \vec{\dot{y}} | \vec{x} )$ needs to
be determined for all subsequences $\dot{\vec{y}}$ within the drift limit
considered.
Observe that the same argument applies equally when the receiver metric is
computed using the lattice implementation \eqref{eqn:F}.
Therefore, when the lattice implementation is used in batch mode, the MAP
decoder has an asymptotic complexity of $\Theta(N n q M_\tau [n + m_n^{+}])$.

\subsection{Optimizing the Lattice Implementation}
\label{sec:lattice_corridor}

In the lattice implementation of the receiver metric, it can be readily seen
that the horizontal distance of a lattice node from the main diagonal is
equivalent to the channel drift for the corresponding transmitted bit.
It should therefore be clear that the likelihood of a path passing through
a lattice node decreases as the distance to the main diagonal increases.

We can take advantage of the above observation by limiting the lattice
computation to paths within a fixed corridor around the main diagonal.
Specifically, the arguments of Section~\ref{sec:statespace} can be applied
directly, resulting in a corridor of width $M_n$ in general for the transmitted
sequence of $n$ bits considered.
Exceptions to this width occur in the first few rows with index $i < -m_n^{-}$
and the last few rows with index $i > \dot\mu - m_n^{+}$, where part of the
corridor falls outside the lattice rectangle.
The number of nodes within this corridor is given by
\begin{align}
   \kappa &= n M_n - \kappa_\mathrm{UL} - \kappa_\mathrm{LR}\text{,} \\
   \text{where~}
   \kappa_\mathrm{UL} &=
      \Delta(-m_n^{-}) - \Delta(-m_n^{-} - n)
      \text{,} \\
   \kappa_\mathrm{LR} &=
      \Delta(n + m_n^{+} - \dot\mu) - \Delta(m_n^{+} - \dot\mu)
      \text{,} \\
   \begin{split}
      \text{and~}
      \Delta(k) &=
      \begin{cases}
         \frac{k^2 + k}{2} & \text{if $k > 0$} \\
         0 & \text{otherwise.} \\
      \end{cases}
   \end{split}
\end{align}
The complexity of the corridor-limited lattice algorithm is therefore
$\Theta(n M_n - \kappa_\mathrm{UL} - \kappa_\mathrm{LR})$.

Some simplification of this expression is possible when the corridor-limited
lattice algorithm is used in the MAP decoder with batch computation for the
channel considered.
When batch computation is used, $\dot\mu = n + m_n^{+}$ by definition,
so that $\kappa_\mathrm{LR} = 0$.
Furthermore, for the BSID channel, $-n \le m_n^{-} \le 0$, so that
$\kappa_\mathrm{UL} = \frac{1}{2}\left[(m_n^{-})^2 - m_n^{-}\right]$.
Under these conditions, the MAP decoder has an asymptotic complexity of
$\Theta\left(N q M_\tau
   \left[n M_n - \frac{1}{2}\left[(m_n^{-})^2 - m_n^{-}\right] \right] \right)$.

\subsection{Comparing Complexity}
\label{sec:complexity_improvement}

A summary of the complexity expressions for the MAP decoder for
various computation modes of the receiver metric is given in
Table~\ref{tab:complexity}.
\inserttab{tab:complexity}{cll}
   {Complexity expressions for the MAP decoder for various computation modes
   of the receiver metric.}{
   & \emph{Algorithm}
      & \emph{Complexity} \\
   \hline
   A & Original
      & $\Theta(N n q M_\tau M_n^2 M_1)$ \\
   B & Batch computation
      & $\Theta(N n q M_\tau M_n M_1)$ \\
   C & Lattice receiver
      & $\Theta(N n q M_\tau [n + m_n^{+}])$ \\
   D & Corridor constraint
      & $\Theta\left(N q M_\tau \left[n M_n -
         \frac{1}{2}\left[(m_n^{-})^2 - m_n^{-}\right] \right] \right)$ \\
   }
%
Comparing the expressions in rows A and B of Table~\ref{tab:complexity}
we can immediately see that the batch computation of the receiver metric
reduces complexity by a factor equal to $M_n$.
Unfortunately, the remaining complexity expressions contain terms that depend
on the code parameters and channel conditions in a rather opaque way, making
it harder to understand the benefits of these improvements.
In the first instance, we can simplify the expressions further to facilitate
comparison.
Consider the expression in row C of Table~\ref{tab:complexity}, when the
lattice implementation is used.
It can be shown that $n + m_n^{+} \rightarrow M_n-1$ as channel conditions
get worse; we can therefore simplify the complexity expression to
$O(N n q M_\tau M_n)$.
Comparing this to the expression in row B of Table~\ref{tab:complexity}
we can see that the use of the lattice implementation reduces complexity
by a factor of at least $M_1$.
Finally, consider the expression in row D of Table~\ref{tab:complexity}, when
the corridor constraint is applied to the lattice algorithm.
Since $m_n^{-} \le 0$, the $\frac{1}{2}\left[(m_n^{-})^2 - m_n^{-}\right]$
term is strictly positive.
The reduction in complexity offered by the corridor constraint is therefore
equal to $\frac{2 n M_n}{2 n M_n - (m_n^{-})^2 + m_n^{-}}$,
and becomes significant as channel conditions improve.
As channel conditions get worse, $m_n^{-} \rightarrow -n$, so that the
expression is dominated by the $n M_n$ term.
Under these conditions, the complexity of the corridor-constrained lattice
implementation becomes approximately equal to that of the unconstrained
lattice implementation.

We can also illustrate the effect of the proposed speedups by considering a
rate-$\frac{1}{2}$ TVB code with typical block and codeword sizes $N=500$,
$n=10$, and $q=32$.
We compute the MAP decoder complexity for this code under a range of channel
conditions, using the original algorithm of Section~\ref{sec:mapdecoder}
and the improvements described above.
We plot these in Fig.~\ref{fig:complexity} using the same summation limits
as in Section~\ref{sec:limits_example}.
\insertfig{fig:complexity}{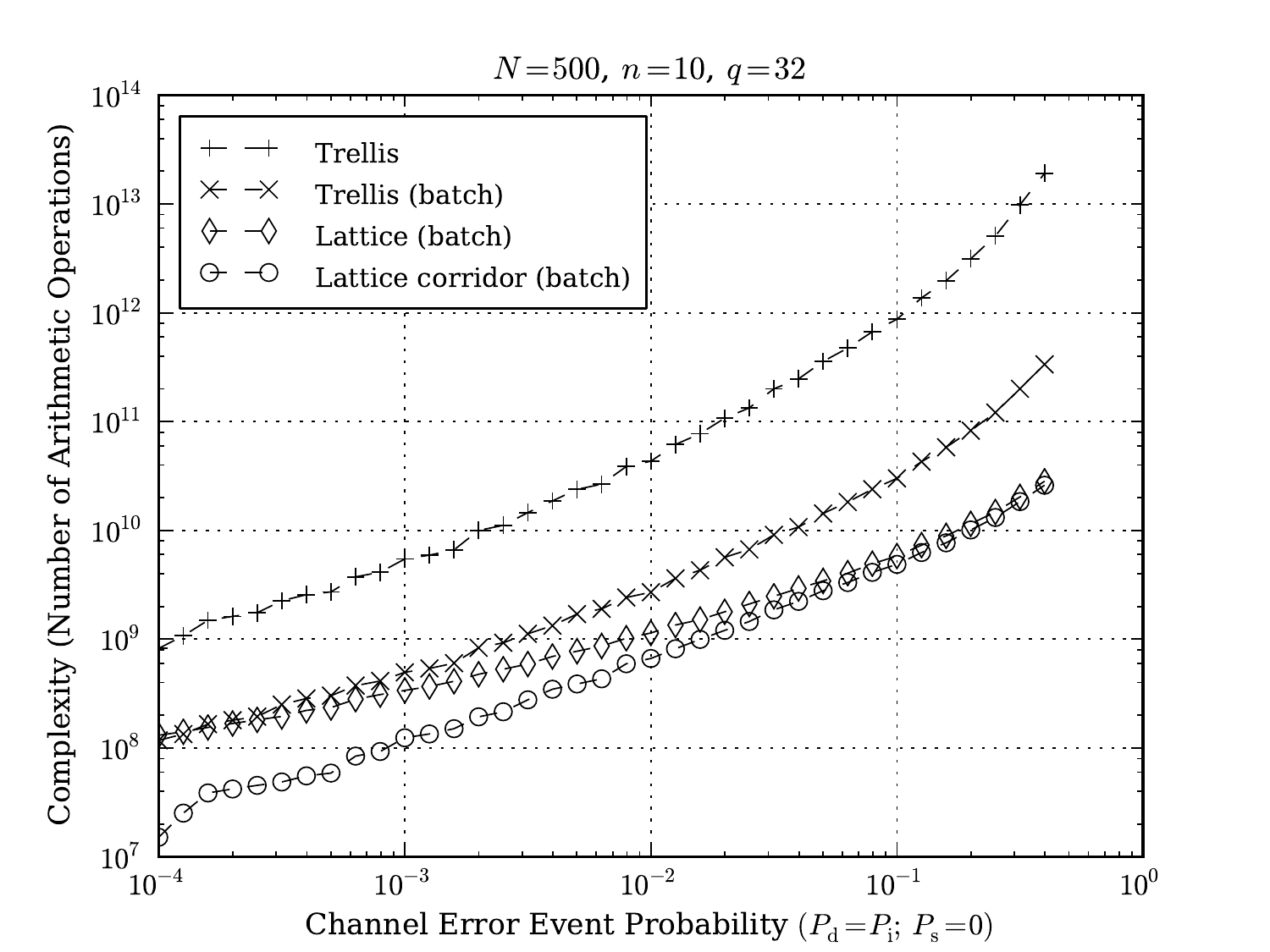}
   {MAP decoder complexity (in number of arithmetic operations) under
   a range of channel conditions, for various computation modes of the
   receiver metric.}
Note that for a fairer comparison between the lattice and trellis modes,
we include a constant factor of three in the lattice computation.
This follows the observation that each lattice node computation
\eqref{eqn:F} requires three multiplications while each trellis computation
\eqref{eqn:alpha_dot} requires only one.
A few general observations can be made on this graph:
\begin{inparaenum}[a)]
\item The batch computation of the receiver metric results in a considerable
reduction of complexity throughout, but is even more significant under poor
channel conditions.
\item The lattice implementation is considerably less complex than the
trellis implementation at high channel error rates.
\item The lattice corridor constraint extends this improvement to the low
channel error rate range.
\end{inparaenum}
In conclusion, the proposed speedups result in a considerable reduction
in complexity of almost two orders of magnitude for typical code sizes and
channel conditions.
We have observed a similar trend under a range of typical code sizes, so
this result can be taken as representative.



\section{Stream Decoding}
\label{sec:streamdecoding}

We have so far considered the case where frame boundaries are known exactly.
While there are practical cases involving single-frame transmission where
this is true, exact frame boundaries are often unknown.
The MAP decoder can handle such cases by changing the initial conditions
for \eqref{eqn:alpha} and \eqref{eqn:beta} and choosing appropriate state
space limits.
This obviates the need for explicit frame-synchronization markers as used
in conventional communication systems, and can therefore reduce this overhead.
The approach presented here is in principle similar to that used in
\cite{dm01ids} for `sliding window' decoding.
However, there are some critical differences which we explore further in
Section~\ref{sec:stream_comparison}.

\subsection{Choosing End-of-Frame Priors}
\label{sec:eof_priors}

Consider first the common case where a sequence of frames is transmitted in
a stream.
The usual practice in communication systems is for the receiver to decode
one frame at a time, starting the decoding process as soon as all the data
related to the current frame is obtained from the channel.
In this case, the current received frame is considered to be
$\vec{Y}\sw{m_\tau^{-}}{\tau+m_\tau^{+}}$,
which may include some bits from the end of the previous frame and start of
the next frame.
The end-state boundary condition for \eqref{eqn:beta} can be obtained by
convolving the expected end-of-frame drift probability distribution with
the start-state distribution:
\begin{equation}
\beta_N(m) = \sum_{m'} \alpha_0(m') \Phi_\tau(m-m')
\text{.}
\end{equation}
Note that in general this distribution $\beta_N(m)$ has a wider spread than
$\Phi_\tau(m)$.

As discussed in Sections~\ref{sec:drift_limits} and \ref{sec:summation_limits},
the choice of state space limits depends on the expected distribution of drift.
For limits involving the whole frame, the distribution used is $\Phi_\tau(m)$,
which assumes that the initial drift is zero.
The assumption does not hold under stream decoding conditions, where the
initial drift is not known \emph{a priori}, although its distribution can
be estimated.
The uncertainty in locating the start-of-frame position increases the
uncertainty in locating the end-of-frame position, resulting in a wider
prior distribution for the end-state boundary condition $\beta_N(m)$.
Therefore, any limits on state space determined using $\Phi_\tau(m)$ will
be underestimated.
The severity of this error depends on the difference between $\beta_N(m)$
and $\Phi_\tau(m)$, which increases as channel conditions get worse.
For stream decoding, therefore, it is sensible to recompute the state space
limit $M_\tau$ at the onset of decoding a given frame, using $\beta_N(m)$
in lieu of $\Phi_\tau(m)$.
Doing so avoids underestimating the required state space, and implies that
for stream decoding, the state space size will change depending on how
well-determined the frame boundaries are.

After decoding the current frame, we obtain the posterior probability
distribution for the drift at end-of-frame, given by:
\ifdraft
   \begin{align}
   \prob{ S_{\tau}=m \;\middle\vert\; \vec{Y}\sw{m_\tau^{-}}{\tau+m_\tau^{+}} }
      & = \lambda_N(m) / \prob{ \vec{Y}\sw{m_\tau^{-}}{\tau+m_\tau^{+}} }
        = \frac{ \lambda_N(m) }{ \sum_{m'} \lambda_N(m') }
      \text{.}
   \end{align}
\else
   \begin{align}
   \prob{ S_{\tau}=m \;\middle\vert\; \vec{Y}\sw{m_\tau^{-}}{\tau+m_\tau^{+}} }
      & = \lambda_N(m) / \prob{ \vec{Y}\sw{m_\tau^{-}}{\tau+m_\tau^{+}} } \nonumber\\
      & = \frac{ \lambda_N(m) }{ \sum_{m'} \lambda_N(m') }
      \text{.}
   \end{align}
\fi
The most likely drift at end-of-frame can be found by:
\begin{align}
\hat S_{\tau}
   & = \argmax_{m} \frac{ \lambda_N(m) }{ \sum_{m'} \lambda_N(m') }
     = \argmax_{m} \lambda_N(m)
   \text{.}
\end{align}
As in \cite{dm01ids}, we determine the nominal start position of the next
frame by shifting the received stream by $\tau + \hat S_{\tau}$ positions.
The initial condition for the forward metric for the next frame,
$\hat\alpha_0(m)$, is set to:
\begin{align}
\hat\alpha_0(m)
   & = \frac{ \lambda_N(m + \hat S_{\tau}) }{ \sum_{m'} \lambda_N(m') }
   \text{,}
\end{align}
replacing the initial condition for \eqref{eqn:alpha} reflecting a known
frame boundary.

\subsection{Stream Look-Ahead}
\label{sec:lookahead}

Taking advantage of the different constituent encodings in TVB codes, the
MAP decoder can make use of information from the following frame to improve
the determination of the end-of-frame position.
We augment the current block of $N$ symbols with the first $\nu$ symbols from
the following block (or blocks, when $\nu > N$), for an augmented block size
$N' = N + \nu$.
The MAP decoder is applied to the corresponding augmented frame.
After decoding, only the posteriors for the initial $N$ symbols are kept;
the start of the next frame is determined from the drift posteriors at the end
of the first $N$ symbols, and the process is repeated.

Consider the latency of the MAP decoder to be the time from when the first
bit of a frame enters the channel to when the decoded frame is available.
The cost of look-ahead is an increase in decoding complexity and latency
corresponding to the change in block size from $N$ to $N'$.
The effect on complexity is seen by using terms corresponding to the augmented
block size in the expressions of Table~\ref{tab:complexity}.
The latency is equal to the time it takes to receive the complete frame and
decode it.
Look-ahead increases the time to receive the augmented frame linearly with
$\nu$ and the decoding time according to the increase in complexity.

The required look-ahead $\nu$ depends on the channel conditions and the
code construction.
In general, a larger value is required as the channel error rate increases.
We show how an appropriate value for $\nu$ can be chosen for a given code
under specific channel conditions in Section~\ref{sec:results_stream}.
Typical values for $\nu$ are small ($\nu < 10$) for good to moderate channels
($\Pri,\Prd < 10^{-2}$).
The required look-ahead increases significantly for poor channels:
the example in Section~\ref{sec:results_stream} requires $\nu=1000$ at
$\Pri=\Prd = 2\times10^{-1}$.

\subsection{Comparison with Davey-MacKay Decoder}
\label{sec:stream_comparison}

A key feature of the Davey-MacKay construction is the presence of a known
distributed marker sequence that is independent of the encoded message.
This allows the decoder, in principle, to compute the forward and backward
passes over the complete stream.
However, to reduce decoding delay, the decoder of \cite{dm01ids} performs
frame-by-frame decoding using a `sliding window' mechanism.
The `sliding window' mechanism seems intended to approximate the computation
of the forward and backward passes over all received data at once.
This approach is similar in principle to ours when stream look-ahead is used;
however, there are some critical differences which we discuss below.

In \cite{dm01ids}, the starting index for a given frame is taken to be the
most likely end position of the previous frame, as determined by the Markov
model posteriors.
This is the same as the approach we use in Section~\ref{sec:eof_priors}.
However, in \cite{dm01ids}, the initial conditions of the forward pass are
simply copied from the final values of the forward pass for the previous frame.
This is consistent with the view that the `sliding window' mechanism
approximates the computation over all received data at once, but contrasts
with our method.
In Section~\ref{sec:eof_priors} the initial conditions of the forward pass
are determined from the posterior probabilities of the drift at the end of
the previous frame.
These drift posteriors include information from the look-ahead region and
from the priors at the end of the augmented frame, which were determined
analytically from the channel parameters.

Observe that in the `sliding window' mechanism of \cite{dm01ids}, the backward
pass values cannot be computed exactly as for the complete stream.
Instead, the decoder of \cite{dm01ids} computes the forward pass for some
distance beyond the expected end of frame position, and initializes the
backward pass from that point.
The suggested distance by which to exceed the expected end of frame position
is `several (e.g.\ five) multiples of $x_\textrm{max}$', where $x_\textrm{max}$
is the largest drift considered.
The concept is the same as the stream look-ahead of
Section~\ref{sec:lookahead}.
However, we recommend choosing the look-ahead quantity $\nu$ based on
empirical evidence (c.f.\ Section~\ref{sec:results_stream}).

It is claimed in \cite{dm01ids} that the backward pass is initialized from the
final forward pass values; the reasoning behind this is unclear, and does not
seem to have a theoretical justification.
We initialize the backward pass with the prior probabilities for the drift
at the end of frame, as explained in Section~\ref{sec:eof_priors}.

\subsection{Initial Synchronization}

The only remaining problem is to determine start-of-frame synchronization
at the onset of decoding a stream.
This can be obtained by choosing state space limits $M_\tau$ large enough to
encompass the initial desynchronization and by setting equiprobable initial
conditions:
$\alpha_0(m) = \beta_N(m) = \frac{1}{M_\tau} \quad \forall m$.
Previous experimental results \cite{dm01ids} have assumed a known start for
the first frame, with the decoder responsible for maintaining synchronization
from that point onwards.
We adopt the same strategy in the following.


\section{Results}
\label{sec:results}

Practical results are given in this section.
We show how an appropriate choice of decoder parameters allows stream decoding
to perform as well as when frame boundaries are known.
Results are also given for existing constructions which can be expressed as
TVB codes, showing how the symbol-level MAP decoder improves on the original
decoder (in the case of \cite{dm01ids}) or is equivalent (in the case of
\cite{ratzer05telecom}).
We also demonstrate some improved constructions allowed by the flexibility
of TVB codes.
These are achieved by using simulated annealing to find TVB codes of a required
order with a good Levenshtein distance spectrum.
Specifically, we seek to find constituent codes with the highest possible
minimum Levenshtein distance and the lowest multiplicity at small distances.
For all codes so designed, $M<N$; we construct our TVB codes using a random
sampling with replacement of the unique constituent codes, and use this as
our inner code.
Construction parameters for all codes used in this section are given in
Table~\ref{tab:codes}.
To facilitate reproduction of these results, the TVB codebooks used are
available for download from the first author's web site\footnote{%
Available at \url{http://jabriffa.wordpress.com/publications/data-sets/}.
}.

\inserttabd{tab:codes}{cllll}
   {Construction parameters of codes used in simulations.}{
   \emph{Label}\footnote{%
   Labels starting with P indicate previously published results, while
   labels starting with N indicate new simulation results.} &
   \emph{Inner code} &
   \emph{Marker} &
   \emph{Outer code} &
   \emph{Comment} \\
   \hline
   P1 &
   $(5,16)$ sparse &
   random, distributed &
   LDPC $(999,888)$ $\mathbb{F}_{16}$ &
   Published in \cite[Fig.~8, Code D]{dm01ids} \\
   N1a &
   $(5,16)$ sparse &
   random, distributed &
   LDPC $(999,888)$ $\mathbb{F}_{16}$ &
   Identical construction to P1, symbol-level MAP decoder \\
   N1b &
   $(10,256,3)$ TVB &
   none &
   LDPC $(499,444)$ $\mathbb{F}_{256}$ &
   Same overall rate and block size as P1 \\
   \hline
   P2 &
   $(6,8)$ sparse &
   random, distributed &
   LDPC $(1000,100)$ $\mathbb{F}_{8}$ &
   Published in \cite[Fig.~8, Code I]{dm01ids} \\
   N2a &
   $(6,8)$ sparse &
   random, distributed &
   LDPC $(1000,100)$ $\mathbb{F}_{8}$ &
   Identical construction to P2, symbol-level MAP decoder \\
   N2b &
   $(6,8,12)$ TVB &
   none &
   LDPC $(1000,100)$ $\mathbb{F}_{8}$ &
   Same overall rate, block size, and outer code as P2 \\
   \hline
   P3 &
   $9$ bits, uncoded &
   $001/110$, appended &
   LDPC $(3001,2000)$ $\mathbb{F}_{2}$ &
   Published in \cite[Fig.~7, Code D]{ratzer05telecom} \\
   N3 &
   $9$ bits, uncoded &
   $001/110$, appended &
   LDPC $(2997,1998)$ $\mathbb{F}_{2}$\footnote{\label{fn:n3a}%
   Obtained by truncating the LDPC $(3001,2000)$ $\mathbb{F}_{2}$ of P3.
   This truncation is necessary so that the outer-encoded sequence can be
   expressed by an integral number of inner codewords.} &
   Identical inner code to P3, marginally smaller outer code \\
   \hline
   P4 &
   not applicable &
   not applicable &
   Rate $\frac{3}{14}$ turbo code $\mathbb{F}_{2}$ &
   Published in \cite[Fig.~4, Code T2]{mans12} \\
   N4 &
   $(7,8,8)$ TVB &
   none &
   LDPC $(666,333)$ $\mathbb{F}_{8}$ &
   Same overall rate as P4 \\
   \hline
   P5 &
   not applicable &
   not applicable &
   Rate $\frac{1}{10}$ turbo code $\mathbb{F}_{2}$ &
   Published in \cite[Fig.~4, Code T4]{mans12} \\
   N5 &
   $(7,4,32)$ TVB &
   none &
   LDPC $(855,300)$ $\mathbb{F}_{4}$ &
   Same overall rate as P5 \\
   }

\subsection{Stream Decoding with MAP Decoder}
\label{sec:results_stream}

The results of \cite{bsw10icc} assumed known frame boundaries; under these
conditions, the decoder is arguably at an advantage when comparing with the
results of \cite{dm01ids, ratzer05telecom}.
It is also not clear whether the MAP decoder can keep track of frame boundaries
with the non-sparse constructions of \cite{bsw10icc, bb11isit} and the TVB
codes introduced here, especially in the absence of a known marker sequence.
In the following we investigate the performance of the MAP decoder under
stream decoding conditions, and consider the choice of look-ahead required.
As in \cite{dm01ids} we assume that the start of the first frame is known,
while the decoder is responsible for keeping synchronization from that point
onwards.
We use the limits specified in Section~\ref{sec:statespace}.

We start by investigating the effect of stream decoding on the ability of the
MAP decoder to track codeword boundaries.
We consider a $(6,8,12)$ TVB code, which is the inner code for the
concatenated system N2b of Table~\ref{tab:codes}.
We simulate this inner code with a block size $N=2000$ under the channel
conditions at the onset of convergence for the concatenated system, that
is at $\Pri = \Prd = 0.22$, assuming only the start position of the first
frame is known.
At each codeword boundary we plot the fraction of correctly determined drifts
(fidelity) in Fig.~\ref{fig:fidelity}.
\insertfig{fig:fidelity}{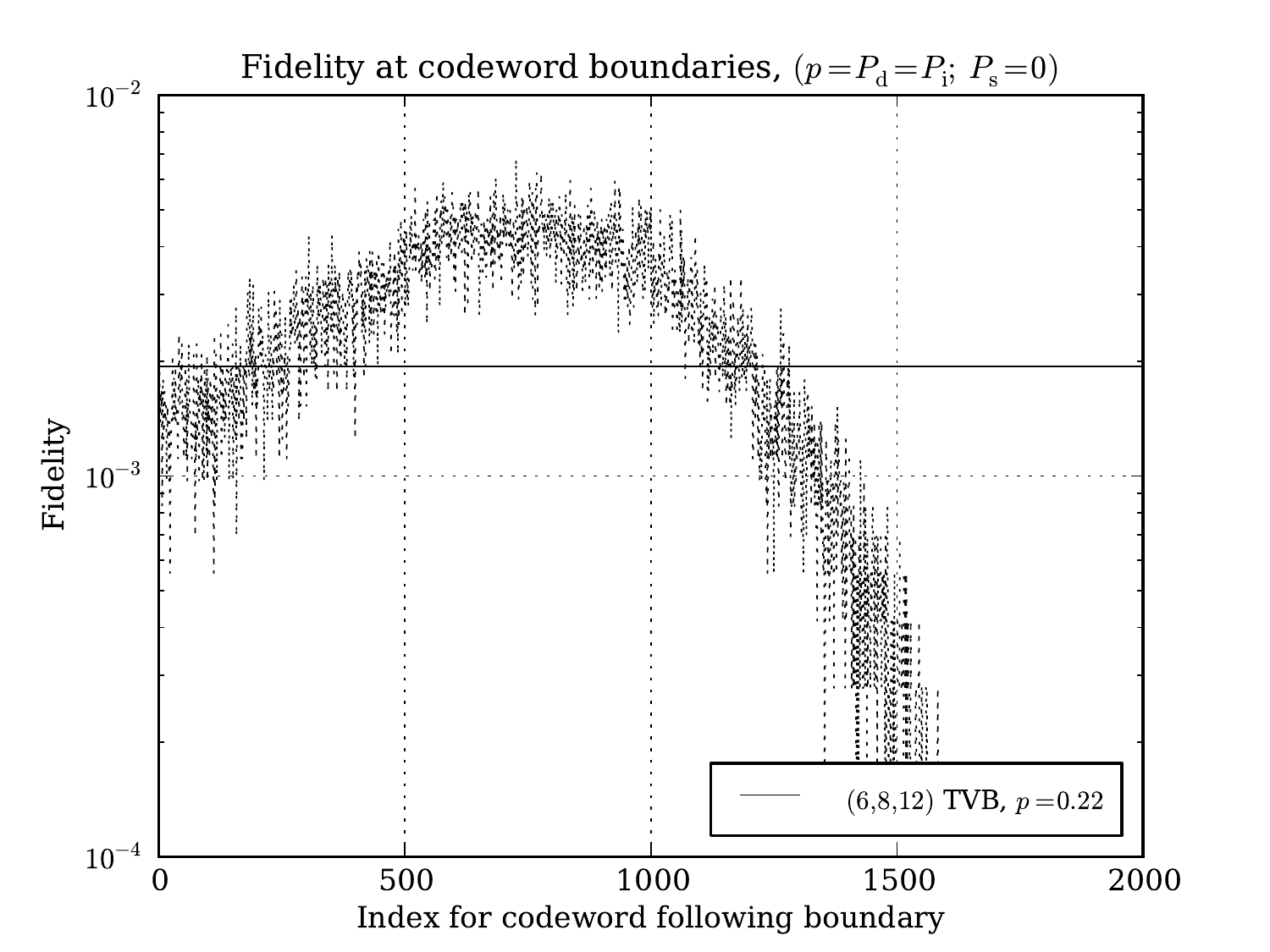}
   {The fraction of correctly resynchronized codeword boundaries (fidelity) as
   a function of codeword index, for code N2b of Table~\ref{tab:codes}.}
As expected, the fidelity drops at the end of the frame, where the actual
drift is unknown to the decoder.
However, it can be observed that the fidelity reaches a steady high value
within about 1000 codewords from the end of frame.
It could therefore be supposed that a look-ahead of $\nu=1000$ would be
sufficient for this code under these channel conditions.
The dip at the start of the frame is caused by the uncertainty in the frame
start position, due to the very low fidelity at the end of the previous frame.

To test this hypothesis, we concatenate this inner code with the $(1000,100)$
LDPC code over $\field{F}_8$ of \cite[Code I]{dm01ids}.
We simulate this system under the following conditions:
\begin{inparaenum}
\item known frame start and end (frame decoding);
\item known start for the first frame, unknown frame ends (stream decoding),
no look-ahead;
\item stream decoding with look-ahead $\nu=1000$ codewords.
\end{inparaenum}
Results are shown in Fig.~\ref{fig:lookahead}.
\insertfig{fig:lookahead}{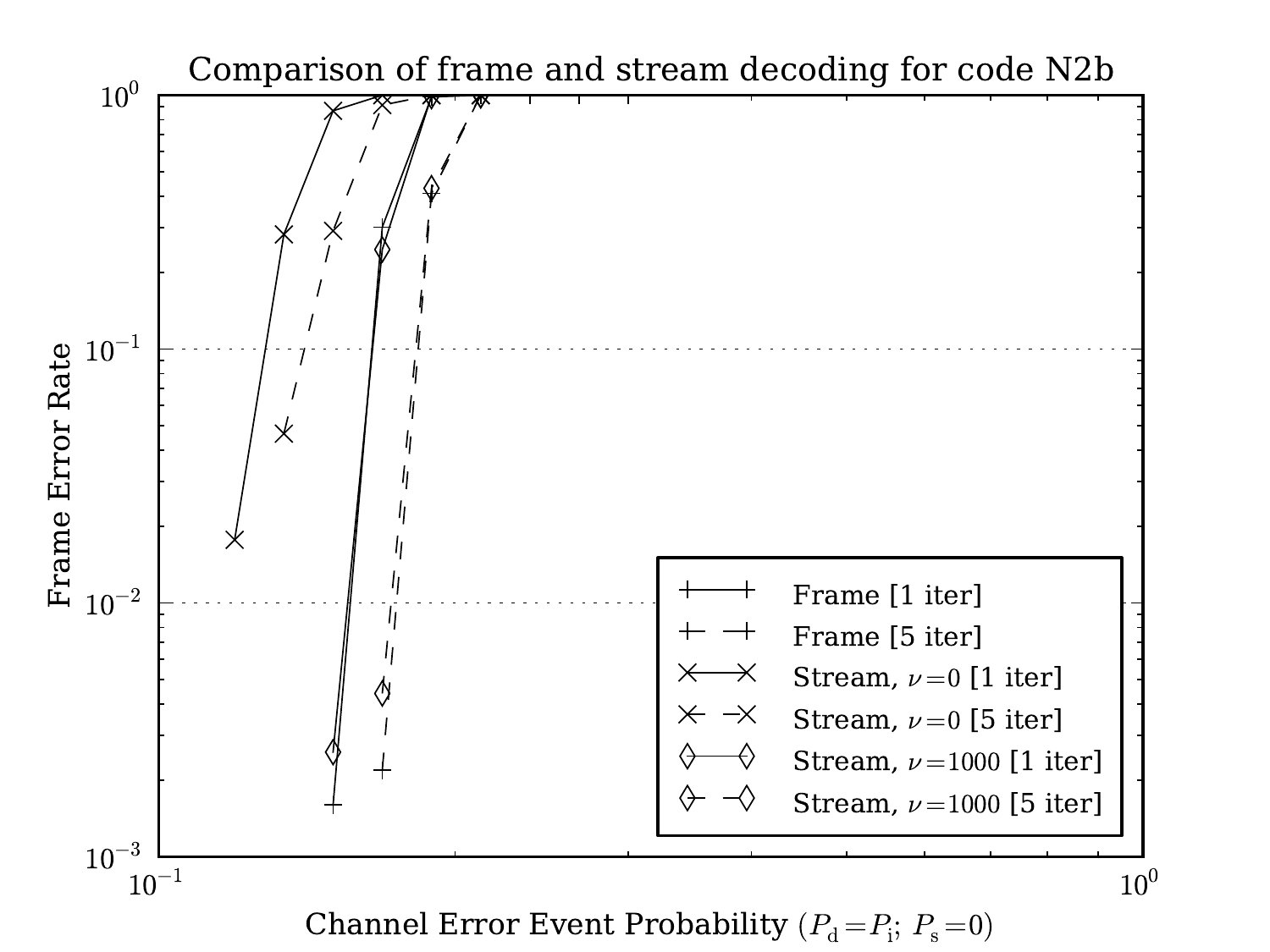}
   {Demonstration of the effect of look-ahead on the MAP decoder's performance
   under stream decoding conditions, and comparison with frame decoding.}
We give results after the first and fifth iterations.
As anticipated, performance under stream decoding conditions is poorer than
frame decoding if there is no look-ahead.
However, an appropriate look-ahead quantity allows the decoder to perform
as well under stream decoding as under frame decoding.

It is important to highlight that this result is dependent on the inner code
structure, and that therefore the generalization to other constructions is
not obvious.
However, we have repeated the same test with other constructions, including
those of \cite{dm01ids, ratzer05telecom, bs08isita, bsw10icc, bb11isit} and
the new constructions in this paper, and under different channel conditions.
In all cases we have found that the result is repeatable, in that it
is possible to approach the performance of frame decoding with stream
decoding, as long as an appropriate look-ahead quantity is chosen.
The only cost of stream decoding is therefore the need for a fidelity analysis
to determine a suitable look-ahead value and the increased decoding latency
and complexity caused by the augmented block size.
Since the code performance is undiminished, to simplify our analysis from
this point onwards we assume known frame start and end positions.

\subsection{Comparison with Prior Art}
\label{sec:results_comparison}


We have already shown in \cite{bsw10icc} that the (symbol-level) MAP decoder
allows us to obtain better performance from the codes of \cite{dm01ids}.
Further improvement can be obtained with iterative decoding, as we show here.
Additionally, the flexibility of TVB codes allows us to obtain codes that
perform better at the same size and/or rate.
In the following, we simulate channel conditions $\Pri=\Prd; \Prs=0$ in
order to compare with published results.

For low channel error rates, consider \cite[Code D]{dm01ids}, listed as P1
in Table~\ref{tab:codes}.
We compare the previously published result with a MAP decoding of the same
code (N1a) in Fig.~\ref{fig:davey}.
\insertfig{fig:davey}{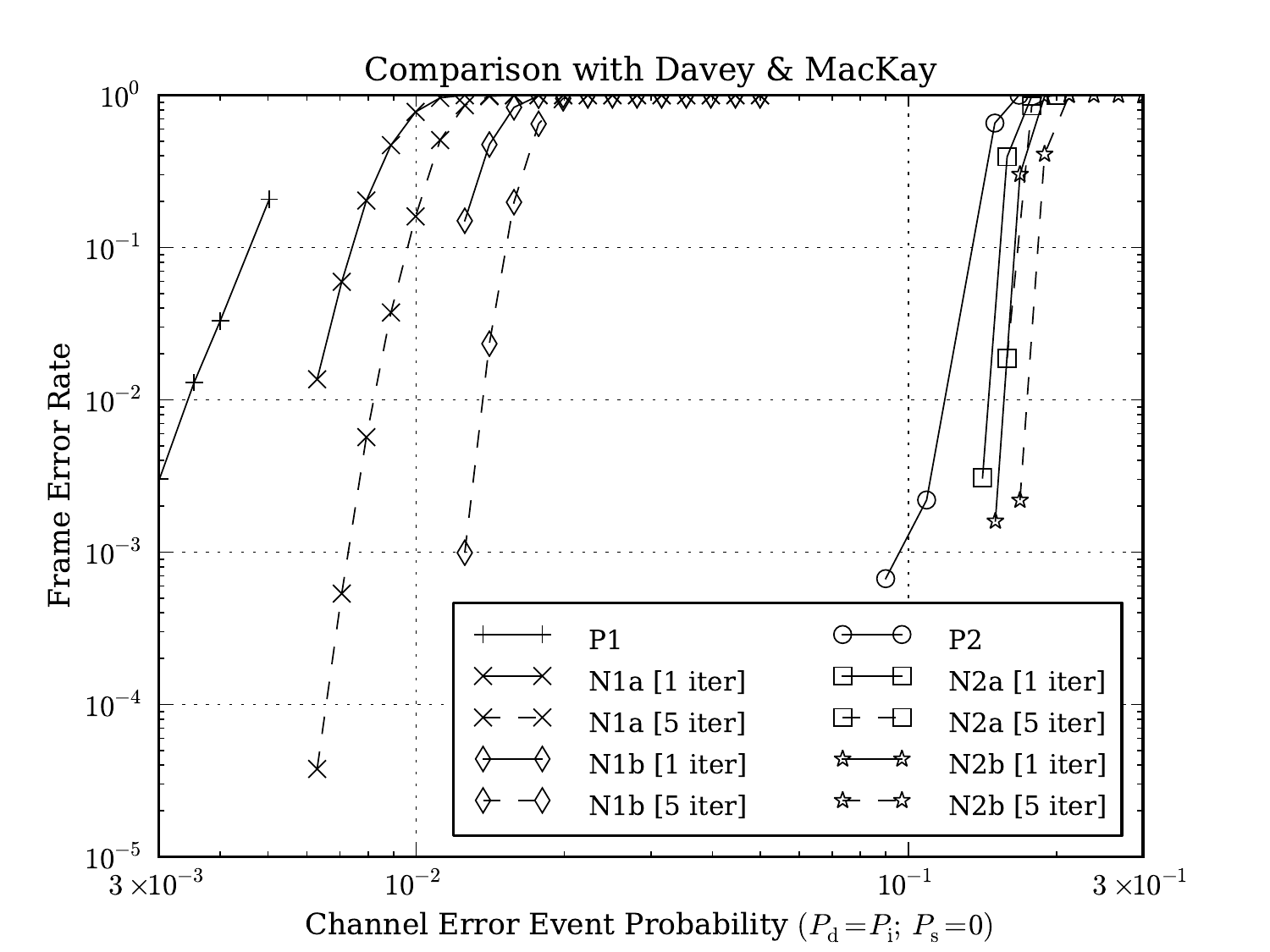}
   {Comparison with Davey-MacKay: improving the performance of
   \cite[Code D]{dm01ids} (left) and \cite[Code I]{dm01ids} (right)
   using our MAP decoder, iterative decoding, and an
   inner code with better Levenshtein distance spectrum.}
As shown in \cite{bsw10icc}, the MAP decoder improves the performance of this
code even after the first iteration; additional iterations improve the result
further.
At the same overall code rate and block size we can improve the performance
further by designing an inner TVB code with a better Levenshtein distance
spectrum (N1b).
We repeat the process at higher channel error rates for \cite[Code
I]{dm01ids}, listed as P2 in Table~\ref{tab:codes}.
Again, compared to the published result, a MAP decoding of the same code
(N2a) improves performance even after the first iteration, also in
Fig.~\ref{fig:davey}.
Additional iterations improve the result, but the difference in this case is
less pronounced.
Replacing the inner code with one of the same size but a better Levenshtein
distance spectrum (N2b) improves performance further.


As we have already discussed in Section~\ref{sec:system_description}, the
marker codes given by Ratzer \cite{ratzer05telecom} can also be cast as TVB
codes.
In \cite{ratzer05telecom} binary outer LDPC codes were used.
To use a binary outer code with our MAP decoder, the bitwise APPs can be
obtained from the $q$-ary symbol APPs by marginalizing over the other bits
\cite[p.~326]{mackay2003}; these are then passed to the decoder for the
binary outer code.
In this case, for a binary outer code we expect the performance of the
concatenated code to be identical, whether the inner code is decoded with the
bit-level (MAP) decoder of \cite{ratzer05telecom} or with our symbol-level
decoder.
We show this in Fig.~\ref{fig:ratzer-mansour} for
\cite[Code~D]{ratzer05telecom}, listed as P3 in Table~\ref{tab:codes}, in
comparison with an almost-identical code (N3) using our MAP decoder.
\insertfig{fig:ratzer-mansour}{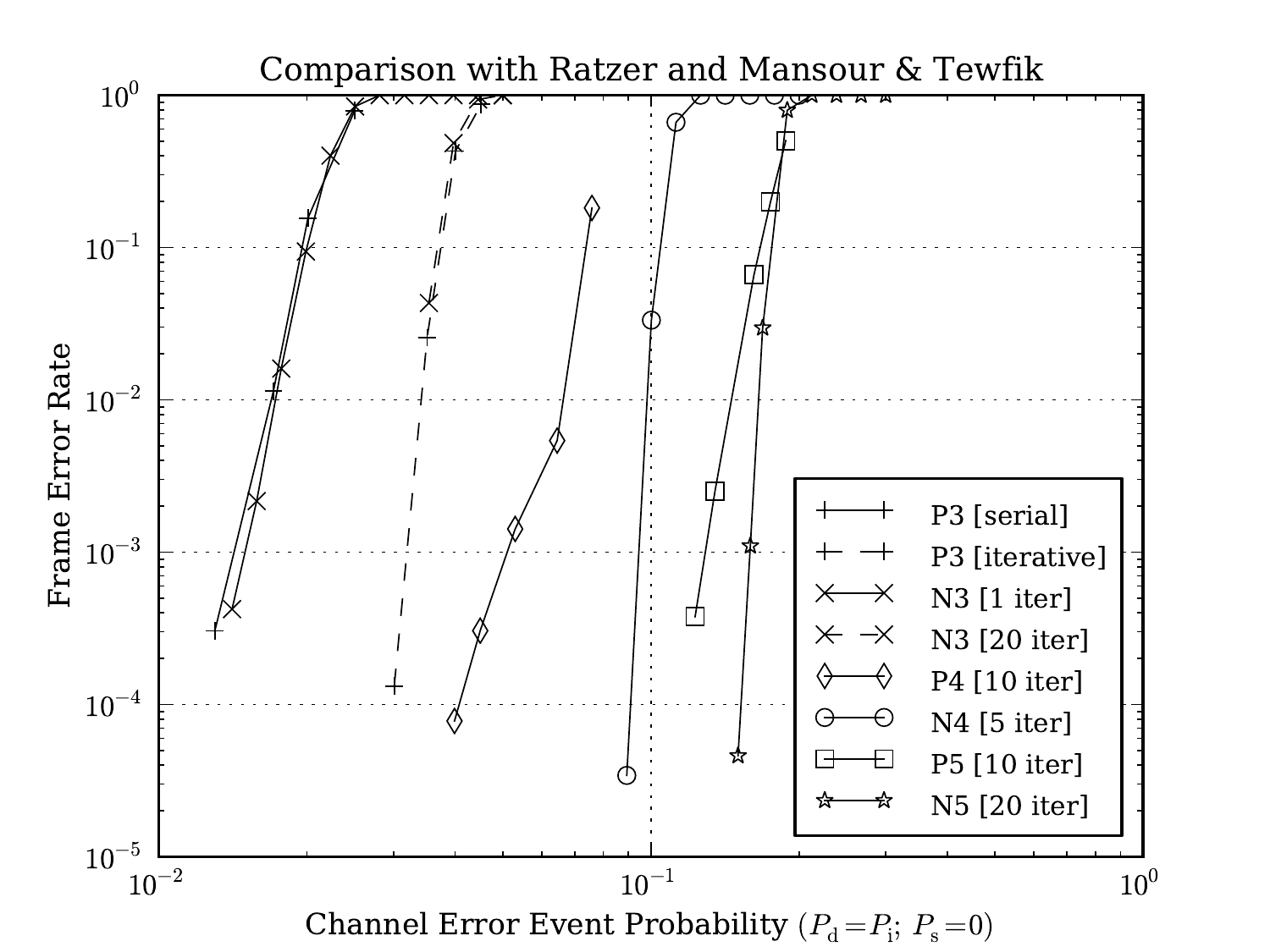}
   {Comparison with Ratzer and with Mansour \& Tewfik:
   demonstrating the equivalence of our MAP decoder on
   \cite[Code~D]{ratzer05telecom}, with and without iterative decoding,
   and improving performance on \cite[Codes T2, T4]{mans12} at the same
   overall code rate, using inner codes with improved Levenshtein distance
   in concatenation with LDPC outer codes.}
It is important to highlight that decoding marker codes as TVB codes provides
no material advantage; in fact, a cost is paid in complexity for doing so.
We do not propose or expect that marker codes will be decoded as TVB codes.
However, there is value in showing that marker codes can be decoded as TVB
codes with no loss in performance.
Specifically, this allows us to compare the structure of marker codes with
other constructions, within the same context.


Finally, in Fig.~\ref{fig:ratzer-mansour} we compare with the more recent
turbo codes of \cite[Codes T2, T4]{mans12}, respectively listed as P4 and P5
in Table~\ref{tab:codes}, with concatenated systems of the same overall rate,
using a TVB inner code and an LDPC outer code.
It can be seen that our concatenated systems outperform the codes of
\cite{mans12}, significantly at lower channel error rates, and somewhat less
so at higher channel error rates.


\section{Conclusions}
\label{sec:closure}

In this paper we have considered TVB codes, which generalize a number of
previous codes for synchronization errors.
We discussed the applicable design criteria for TVB codes, expressing
some previously published codes as TVB codes and showing that the greater
flexibility of TVB codes allows improved constructions.
For example, our $(7,8,4)$ TVB code achieves a SER of $10^{-4}$ at a
$\Pri,\Prd$ that is almost two orders of magnitude higher than a marker
or distributed marker code of the same size, and slightly better than our
earlier SEC codes.

We also considered a number of important issues related to a practical
implementation of the corresponding MAP decoder.
Specifically, we have given an expression for the expected distribution of
drift between transmitter and receiver due to synchronization errors,
with consideration for practical concerns when evaluating this expression.
We have shown how to determine an appropriate choice for state space limits
based on the drift probability distribution.
The decoder complexity under given channel conditions is then expressed as
a function of the state space limits used.
For a given state space, we have also given a number of optimizations that
reduce the algorithm complexity with no further loss of decoder performance.
The proposed speedups, which are independent of the TVB code construction,
result in a considerable reduction in complexity of almost two orders of
magnitude for typical code sizes and channel conditions.
For code constructions with appropriate mathematical structure we expect to be
able to replace the receiver metric, which considers each possible transmitted
codeword, with a faster soft-output algorithm.
Next, we have considered the practical problem of stream decoding, where
there is no prior knowledge of the received frame boundary positions.
In doing so we have also shown how an appropriate choice of decoder parameters
allows stream decoding to approach the performance when frame boundaries
are known, at the expense of some increase in complexity.

Finally, practical comparisons of TVB codes with earlier constructions were
given, showing that TVB code designs can in fact achieve improved performance.
Even compared to the state of the art codes of \cite{mans12}, the TVB
codes presented here achieve a FER of $10^{-3}$ at 24\% higher $\Pri,\Prd$
for a rate-$\frac{1}{10}$ code, and at 84\% higher $\Pri,\Prd$ for a
rate-$\frac{3}{14}$ code.
We expect further improvements to the codes shown here to be possible,
particularly by co-designing optimized outer codes.
However, a detailed treatment of the design process is beyond the scope of this
paper, and will be the subject of further work.


\balance

\bibliographystyle{IEEEtran}
\bibliography{IEEEabrv,jbabrv,jb,turbo,unsorted,victor}

\end{document}